\documentclass[prd,nofootinbib,preprintnumbers]{revtex4-2}
\usepackage{subfig}
\usepackage{amsmath}
\usepackage{appendix}
\usepackage{amsfonts}
\usepackage{caption}
\usepackage{amssymb}
\usepackage{graphicx}
\usepackage{slashed}
\usepackage{orcidlink}
\usepackage[parfill]{parskip}
\usepackage{nicefrac}
\usepackage{mathtools}
\usepackage[normalem]{ulem}

\usepackage{comments}

\newcommand{\xBj}{x_{\text{Bj}}}

\begin{document}

\title{Exclusive Quark and Gluon Dijet Production as Probes of GPDs at Collider Energies}

\author{
Zhuoyi~Pang$^{1}$\orcidlink{0000-0002-8581-8498},
Paweł~Sznajder$^{1}$\orcidlink{0000-0002-2684-803X},
Lech~Szymanowski$^{1}$\orcidlink{0000-0003-1623-8331},
Jakub~Wagner$^{1}$\orcidlink{0000-0001-8335-7096}
}

\affiliation{
$^1$ National Centre for Nuclear Research, NCBJ, 02-093 Warsaw, Poland 
}

\begin{abstract}
{We study exclusive electroproduction of dijets in the collinear factorization framework as a probe of generalized parton distributions (GPDs). For quark dijet production, we extend previous analyses by including contributions from helicity GPDs and by assessing an additional leading-order electromagnetic channel governed by elastic nucleon form factors. Furthermore, we investigate exclusive gluon dijet production. We compare our prediction to HERA data and provide projections for measurements at the future Electron-Ion Collider.}

\end{abstract}
\maketitle

\section{Introduction}
\label{sec:intro}
Introduced three decades ago ~\cite{Muller:1994ses,Ji:1996nm,Radyushkin:1997ki}, generalized parton distributions (GPDs) have become
powerful theoretical tools encoding the three-dimensional structure of hadrons ~\cite{Diehl:2003ny,Belitsky:2005qn}. They also provide access to fundamental quantities such as spin~\cite{Ji:1996ek}, mass~\cite{Ji:1994av}, and internal force distributions of hadrons~\cite{Polyakov:2002yz}. Experimentally, GPDs can be accessed through hard exclusive processes within the collinear factorization framework~\cite{Collins:1996fb,Collins:1998be}. Their study is among the primary goals of the ongoing 12 GeV program at Jefferson Lab \cite{Dudek:2012vr}, as well as the future Electron-Ion Collider (EIC)~\cite{Accardi:2012qut,AbdulKhalek:2021gbh} and Electron-ion collider in China (EicC)~\cite{Anderle:2021wcy}. 

A broad phenomenological program has been developed to interpret hard exclusive scattering data, including both model dependent approaches and more flexible data driven parameterizations such as those based on artificial neural networks  \cite{Kumericki:2009uq, Kroll:2012sm, Moutarde:2018kwr, Kumericki:2016ehc, Moutarde:2019tqa,Cuic:2020iwt}. Complementing these efforts, direct lattice QCD calculations of GPDs, which are based on
large-momentum effective theory (LaMET)~\cite{Ji:2013dva,Ji:2014gla,Ji:2020ect} and short-distance factorization (SDF)~\cite{Radyushkin:2017cyf}, have advanced considerably in recent years~\cite{Alexandrou:2020zbe,Lin:2020rxa,Bhattacharya:2022aob,Chu:2025kew}. 
Although at a preliminary stage, the existing experimental and lattice data have provided valuable guidance on the extractions of GPDs~\cite{Cichy:2024afd,Chu:2025jsi,Guo:2025muf}.

In this work, we consider the exclusive electroproduction of quark and gluon dijets. At LO in $\alpha_s$, the produced quark and gluon dijets can be approximated as a quark-antiquark pair and a gluon-gluon pair:
\begin{align}
\label{eq:reactioneToqq}
&e(l)\,N(p)\rightarrow e(l^{\prime})\,N(p^{\prime})\,q(q_1)\bar{q}\,(q_2) \,, \\
\label{eq:reactioneTogg}
&e(l)\,N(p)\rightarrow e(l^{\prime})\,N(p^{\prime})\,g(q_1)\,g(q_2) \,,
\end{align}
where the symbols in parentheses denote the four-momenta of the respective particles, and we define $q=l-l^\prime.$ This process was first measured by the ZEUS collaboration, using the $k_T$-cluster algorithm for jet reconstruction~\cite{ZEUS:2015sns}. One of the original motivations for studying ~\eqref{eq:reactioneToqq} was to explore the nature of the Pomeron exchanged between the virtual photon and the proton at HERA energies, where the colliding energy is much larger than other hard scales in the process. To this end, various descriptions of the differential cross section have been put forward based on $k_t$ factorization or the small-x formalism~\cite{Nikolaev:1994cd,Bartels:1996tc,Bartels:1996ne,Boussarie:2019ero}, and more recently in terms of GTMD framework~\cite{Linek:2024dzs,Boer:2023mip,Boer:2021upt}, including the exclusive electroproduction of open heavy-quark pairs \cite{Bellotti:2026bau}.

In addition, the existence of (hard) dijet transverse momentum makes ~\eqref{eq:reactioneToqq} a potentially promising process for measuring gluon Wigner distribution and gluon orbital angular momentum (OAM), through certain correlations between the transverse momenta of the dijet and the outgoing nucleon~\cite{Hatta:2016dxp,Ji:2016jgn}. For other observables capable of probing gluon OAM through ~\eqref{eq:reactioneToqq}, see Refs~\cite{Bhattacharya:2022vvo,Bhattacharya:2024sck}.

The focus of this paper is to explore the possibility of probing GPDs through the exclusive electroproduction of dijets, within the collinear factorization framework. The first work in this direction was done by Braun and Ivanov~\cite{Braun:2005rg}, who considered the quark dijet production. Taking into account the contributions from unpolarized GPDs they provided predictions for various differential cross sections within HERA kinematics. In a recent work~\cite{Chall:2026oes}, more detailed numerical analyses were performed, taking into account effects from experimental cuts of H1 and ZEUS experiments. In our work, we will derive the leading-order expressions for processes ~\eqref{eq:reactioneToqq} and ~\eqref{eq:reactioneTogg}, for the amplitude and cross section level. In addition to the new results for gluon dijet production (Eq.~\eqref{eq:reactioneTogg}) and contributions from helicity GPDs to process ~\eqref{eq:reactioneToqq}, we will also provide the discussions about the previously neglected QED channel, which is analogous to the Bethe-Heitler channel in the double deeply virtual Compton scattering (DDVCS).
We then present our numerical results at EIC and HERA kinematics, implemented within the PARTONS \cite{Berthou:2015oaw} framework.

The remainder of this paper is organized as follows. In Sec. ~\ref{sec:kinematics}, the kinematics of exclusive electroproduction of dijets is introduced. In Sec. ~\ref{scattering amp}, we first review the definitions of GPDs, and then give the leading-order expressions for Eq.~\eqref{eq:reactioneToqq} and Eq.~\eqref{eq:reactioneTogg}, at amplitude level. Sec. ~\ref{diff cross section} is devoted to the analytical results for differential cross sections.
In Sec. ~\ref{pheno},
we present detailed phenomenological analyses 
at EIC and HERA kinematics. Finally, we draw our conclusions in Sec. ~\ref{con}.

\section{Kinematics}
\label{sec:kinematics}

In this section, we fix the notation and introduce the kinematic variables used throughout the text. We start with the usual variables used in inclusive deep inelastic scattering (DIS), 
\begin{equation} \label{variable}
y=\frac{q\cdot p}{l\cdot p}\,,
\quad 
Q^2=-q^2\,,
\quad 
W^2=(p+q)^2\,,
\quad 
\xBj=\frac{Q^2}{2p\cdot q}\approx\frac{Q^2}{Q^2+W^2}\,.
\end{equation}
Here, $y$ is the so-called inelasticity variable, $Q^2$ denotes the photon virtuality for the QCD channel considered below, $W$ is the center-of-mass energy of the $\gamma^*\,N$ system, and $\xBj$ is the Bjorken variable. The proton mass will be ignored in the following calculations (which gives $``\approx"$ in the last formula).  The invariant mass of the produced dijet will be denoted by
\begin{equation}
    M^2=(q_1+q_2)^2 \,.
\end{equation}
We also introduce two new four-momenta describing the kinematics of the nucleon, and the associated Mandelstam variable $t$ describing the transfer of four-momentum,
\begin{equation}
P=\frac{p^{\prime}+p}{2}\,,\quad
\Delta=p^{\prime}-p\,,\quad 
t=\Delta^2\,.
\end{equation}
The set of variables is complemented by $\phi$ and $\phi_{p'}$, which denote the azimuthal angles of the produced quark and the produced proton, see Eq.~\eqref{angles}.

The analytic calculations presented in this manuscript utilize the decomposition of four-momenta in the Sudakov basis. For any four-momentum $a$ one has\footnote{Note that $\mathbf{a_\perp^2}=-a_\perp^2$ is always positive.}
\begin{align}
a^\mu=a^+ n_+^\mu + a^- n_-^\mu + a_{\perp}^{\mu}\,,\quad
a^2 = 2a^+a^- -\mathbf{a_{\perp}^2}\,,\quad
\hat{a}=(a^+,a^-,\mathbf{a_\perp})
\end{align}
where $n_+$ and $n_-$ are two light-like vectors,
satisfying 
\begin{equation} 
n_-^2=n_+^2=0\,,\qquad 
n_-\cdot n_+=1\,,\quad
n_- \cdot a_{\perp} = n_+ \cdot a_{\perp} =0\,.
\end{equation}
The vector $n_+$ ($n_-$) has a positive (negative) $z$ component.
We complement our definitions with the transverse metric $g_\perp^{\mu\nu}\equiv g^{\mu\nu}-n_-^\mu n_+^\nu-n_-^\nu n_+^\mu$ and the transverse Levi-Civita tensor $\epsilon_\perp^{\mu\nu}\equiv \epsilon^{n_-n_+\mu\nu}\equiv \epsilon^{\alpha\beta\mu\nu}n_{-\alpha}n_{+\beta}$, where $\epsilon^{0123}=1$.

\begin{figure}[htbp]
\centering
\includegraphics[width=0.6\textwidth]{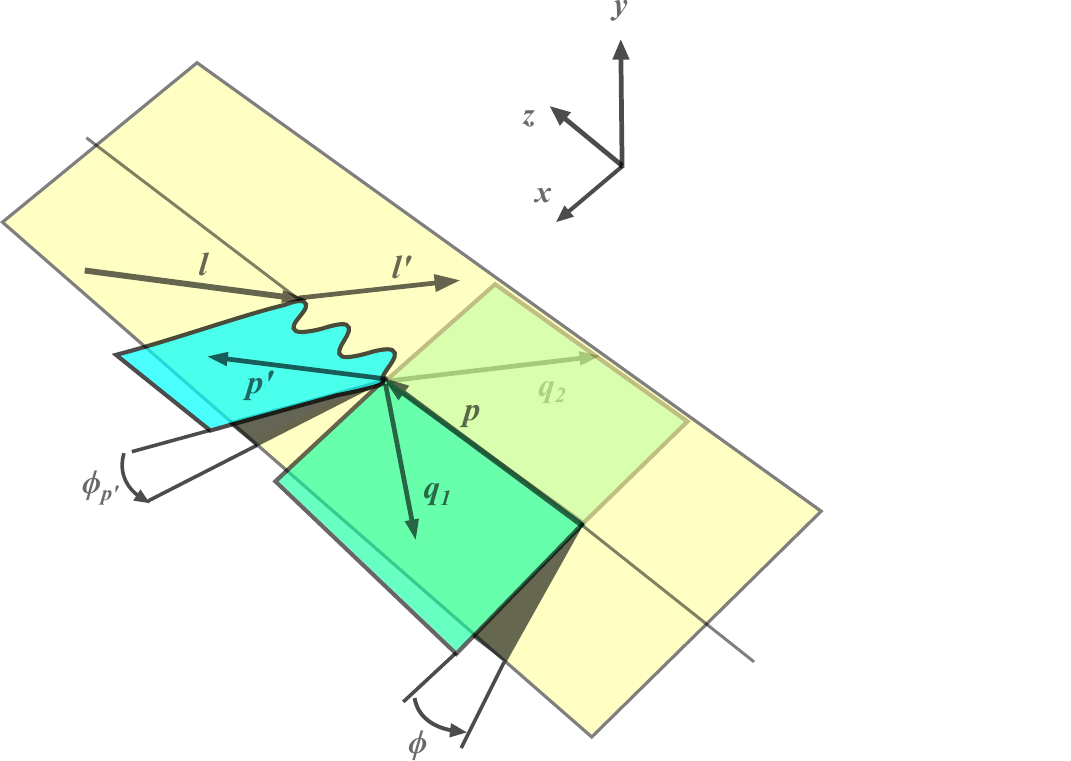}
\caption{Kinematics of exclusive dijet production.}
\label{kinematic figure}
\end{figure}

The calculations are performed in the reference frame where the momentum $q$ and $p$ are back-to-back, see Fig.~\ref{kinematic figure} as an illustration of the QCD channel (defined in Sec. ~\ref{scattering amp}), and we neglect the nucleon mass whenever possible. We introduce the skewness parameter,
\begin{equation}
    \xi=\frac{p^+-p^{\prime+}}{p^++p^{\prime+}} =
    \frac{\sqrt{(Q^2 + M^2)^2 + t^2 + 2t(Q^2 - M^2)}}{2\displaystyle\frac{Q^2}{\xBj} - Q^2 - M^2 + t}
    \,,
\end{equation}
and express the four-momenta in the basis of the chosen light-cone vectors:
\begin{align} \label{setting1}
l^{\mu}=& \frac{(1-y)x_{\text{Bj}}(1+\xi)W}{y} n_+^\mu+\frac{Q^2+W^2}{2W(1+\xi)y}n_-^\mu+l_{\perp}^\mu\,,\nonumber \\
l^{\prime\mu}=&\frac{x_{\text{Bj}}(1+\xi)W}{y}n_+^\mu+\frac{(Q^2+W^2)(1-y)}{2W(1+\xi)y}n_-^\mu+l_{\perp}^\mu\,,\nonumber\\
p^{\mu}=&W(1+\xi) n_+^\mu,\nonumber \\
p^{\prime\mu}=&W(1-\xi) n_+^\mu+\frac{\mathbf{\Delta_\perp^2}}{2(1-\xi)W}n_-^\mu+\Delta_\perp^\mu\,,\nonumber \\
q^{\mu}=&-W(1+\xi) \xBj n_+^\mu+\frac{Q^2+W^2}{2W(1+\xi)}n_-^\mu\,. 
\end{align}
In the collinear factorization framework, at leading-twist, $\Delta_\perp$ is neglected
in the hard part. Under these conditions, we have:  
\begin{align}
q_1^\mu=&\frac{\mathbf{\mathbf{q_\perp^2}}+m_q^2}{z Q^2}W(1+\xi) \xBj n_+^{\mu}+z\frac{Q^2+W^2}{2W(1+\xi)}n_-^{\mu}+q_\perp^\mu\,,\nonumber \\
q_2^\mu=&\frac{\mathbf{q_\perp^2}+m_q^2}{\bar{z} Q^2}W(1+\xi) \xBj n_+^{\mu}+\bar{z}\frac{Q^2+W^2}{2W(1+\xi)}n_-^{\mu}-q_\perp^\mu,
\end{align}
where $z$ is the light-cone minus-momentum fraction carried by one of the partons in the jets (specifically the quark in reaction Eq.~\eqref{eq:reactioneToqq}), $\bar{z} = 1-z$ is the fraction carried by the second parton, and $m_q$ is the quark mass, which is kept only for charm and bottom quark dijets. We define $\big|q_\perp\big|\equiv\sqrt{\mathbf{q_\perp^2}}$ for later convenience. The azimuthal angles $\phi$ and $\phi_{p^\prime}$ are defined as:
\begin{equation} \label{angles}
\text{cos}\phi=\frac{q_\perp^x}{\big|q_\perp\big|},\quad \text{cos}\phi_{p^\prime}=\frac{\Delta_\perp^x}{\sqrt{\mathbf{\Delta_\perp^2}}}.
\end{equation}

In the calculation of the amplitudes, we will frequently use the following relations:
\begin{equation}\label{rela}
\frac{2\xi}{1+\xi}=\frac{Q^2+M^2}{Q^2+W^2}\,,\quad
M^2=\frac{\mathbf{q_\perp^2}+m_q^2}{z\bar{z}}\,.
\end{equation} 
In addition, we introduce the auxiliary variables:
\begin{equation} \label{betamu}
\mu^2=m_q^2+z\bar{z}Q^2\,, \quad
\beta=\frac{\mu^2}{\mathbf{q_\perp^2}+\mu^2} =\left(Q^2+\frac{m_q^2}{z\bar{z}}\right)/\left(M^2+Q^2\right)\,.
\end{equation}
Note that for light quark and gluon dijets, the $\beta$ in Eq.~\eqref{betamu} coincides with the conventional $\beta$ parameter used in the description of diffractive dijet production (which we denote by $\beta^\prime$):
\begin{equation} \label{betadiff}
\beta^\prime=\frac{Q^2}{2q\cdot(p-p^\prime)}=\frac{Q^2}{Q^2+M^2-t}\approx\frac{Q^2}{Q^2+M^2}.
\end{equation}
We have used the fact that $Q^2,M^2\gg t$.

\section{The scattering amplitudes}
\label{scattering amp}

\subsection{Nonperturbative quantities}
In the collinear factorization, the nonperturbative parts can be parametrized using generalized parton distributions (GPDs) and elastic form factors (EFFs). The leading-twist GPDs are defined in ~\cite{Diehl:2003ny,Belitsky:2005qn}. In this paper, we consider the case when the target is unpolarized. 

For the leading-twist quark GPDs, we have:
\begin{align} \label{quark_GPD}
F_{qu}(x,\xi,t)=&\frac{1}{2}\int\frac{\mathrm{d}z^-}{2\pi}\operatorname{e}^{\mathrm{i}xP^+z^-}\langle p^{\prime}|\bar{q}(-\frac{1}{2}z)\gamma^+W(-\frac{1}{2}z,\frac{1}{2}z)q(\frac{1}{2}z)|p\rangle\big|_{z^+=0,z_\perp=0} \nonumber \\
=&\frac{1}{2P^{+}}\Big[H^{q}(x,\xi,t)\bar{u}(p^{\prime})\gamma^{+}u(p)+E^{q}(x,\xi,t)\bar{u}(p^{\prime})\frac{\mathrm{i}\sigma^{+\alpha}\Delta_{\alpha}}{2m_p}u(p)\Big],\nonumber \\
F_{qh}(x,\xi,t)=&\frac{1}{2}\int\frac{\mathrm{d}z^-}{2\pi}\operatorname{e}^{\mathrm{i}xP^+z^-}\langle p^{\prime}|\bar{q}(-\frac{1}{2}z)\gamma^+\gamma_5W(-\frac{1}{2}z,\frac{1}{2}z)q(\frac{1}{2}z)|p\rangle\big|_{z^+=0,z_\perp=0}\nonumber \\
=&\frac{1}{2P^{+}}\Big[\tilde{H}^{q}(x,\xi,t)\bar{u}(p^{\prime})\gamma^{+}\gamma_{5}u(p)+\tilde{E}^{q}(x,\xi,t)\bar{u}(p^{\prime})\frac{\gamma_{5}\Delta^{+}}{2m_p}u(p)\Big].
\end{align}
For the leading-twist gluon GPDs, we have:
\begin{align} \label{gluon_GPD}
F_{gu}(x,\xi,t)=&\frac{1}{P^{+}}\int\frac{\mathrm{d}z^{-}}{2\pi}\mathrm{e}^{\mathrm{i}xP^{+}z^{-}}\langle p^{\prime}|G^{+\mu}(-\frac{1}{2}z)W(-\frac{1}{2}z,\frac{1}{2}z)G_{\mu}^{+}(\frac{1}{2}z)|p\rangle\big|_{z^+=0,z_\perp=0}\nonumber \\
=&\frac{1}{2P^{+}}\Big[H^{g}(x,\xi,t)\bar{u}(p^{\prime})\gamma^{+}u(p)+E^{g}(x,\xi,t)\bar{u}(p^{\prime})\frac{\mathrm{i}\sigma^{+\alpha}\Delta_{\alpha}}{2m_p}u(p)\Big],\nonumber \\
F_{gh}(x,\xi,t)=&-\frac{\mathrm{i}}{P^{+}} \int \frac{\mathrm{d}z^{-}}{2\pi} \mathrm{e}^{\mathrm{i}xP^{+}z^{-}} \langle p^{\prime}|G^{+\mu}(-\frac{1}{2}z)W(-\frac{1}{2}z,\frac{1}{2}z)\tilde{G}_{\mu} ^{+}(\frac{1}{2}z)| p\rangle\big|_{z^+=0,z_\perp=0}\nonumber \\
=&\frac{1}{2P^{+}}\Big[\tilde{H}^{g}(x,\xi,t)\bar{u}(p^\prime)\gamma^{+}\gamma_{5}u(p)+\tilde{E}^{g}(x,\xi,t)\bar{u}(p^\prime)\frac{\gamma_{5}\Delta^{+}}{2m_p}u(p)\Big].
\end{align}
$m_p$ denotes the proton mass, $W(-z^-/2,z^-/2)$ denotes the gauge link connecting fields along light-cone minus direction. The subscripts ``u" and ``h"
denote ``unpolarized" and ``helicity" respectively. We have $\sigma^{\mu\nu}=i\big[\gamma^\mu,\gamma^\nu\big]/2$, $\tilde{G}^{+\mu}= G^{+}_\nu\epsilon_\perp^{\mu\nu}/2.$ For gluon dijet production, there is no contribution from 
transversity GPDs due to conservation of helicity in the hard part. For quark dijet production,
we neglect the contributions from transversity GPDs due to the following reason: For the contributions from gluon transversity GPDs to the scattering amplitude, they vanish in the forward limit and are proportional to the $(\sqrt{t_0-t})^i(i\geq 1)$ factor~\cite{Diehl:2001pm}, which results in further suppression compared to GPD $H_g$ and GPD $\tilde{H}_g$.

The EFFs have the following standard form:
\begin{align} \label{EFF}
J^\alpha_f=&\big<p^\prime\big|\bar{\psi}_f(0)\gamma^\alpha\psi_f(0)\big|p\big>\nonumber \\
=& \bar{u}(p^\prime)\Big[F_1^f(t)\gamma^\alpha+F_2^f(t)\frac{i\sigma^{\alpha\beta}\Delta_\beta}{2m}\Big]u(p),
\end{align}
where $F_1^f(t)$ and $F_2^f(t)$ are Dirac and Pauli FFs for flavor $f$.

\subsection{Quark dijet production}
For quark dijet production (shown in Fig.~\ref{quark_dijetq}--Fig.~\ref{primakofffig}), the amplitude receives contributions not only from quark and gluon GPDs (denoted by $M^{q\bar{q}}_{\text{QCD}}$, called by ``QCD channel" below), but also from EFFs (denoted by $M^{q\bar{q}}_{\text{QED}}$, called by ``QED channel" below), in the latter case a photon is exchanged in the t-channel.

For contributions from quark and gluon GPDs, we can write the amplitude as:
\begin{equation} \label{amplitude}
M^{q\bar{q}}_{\text{QCD}}=\frac{\sqrt{4\pi\alpha_{\text{em}}}}{Q^2}\bar{u}(l^\prime)\gamma^\mu u(l) T^\nu g_{\mu\nu},
\end{equation}
where the superscript ``~$q\bar{q}$~" denotes the quark dijet production. $T^\nu$ denotes the amplitude for the subprocess: $$\gamma^*(q)N(p)\rightarrow N(p^\prime)q(q_1)\bar{q}(q_2).$$
From gauge invariance ~\cite{Braun:2005rg}, $g^{\mu\nu}$ in Eq.~\eqref{amplitude} can be replaced by the polarization sum $\epsilon_L^\mu\epsilon_L^\nu-e_x^\mu e_x^\nu-e_y^\mu e_y^\nu$ in our calculations, where
\begin{align} \label{Ward}
\epsilon_L^\mu=&\frac{W(1+\xi)x_{\text{Bj}}}{Q}n_+^\mu+\frac{Q^2+W^2}{2WQ(1+\xi)}n_-^\mu,\nonumber \\
e_x^\mu=&(0,0,1,0),\quad e_y^\mu=(0,0,0,1).
\end{align}

\subsubsection{Contribution from quark GPDs} \label{QCD_channel_ampq}
\begin{figure}[htbp]
\centering
\includegraphics[width=0.8\textwidth]{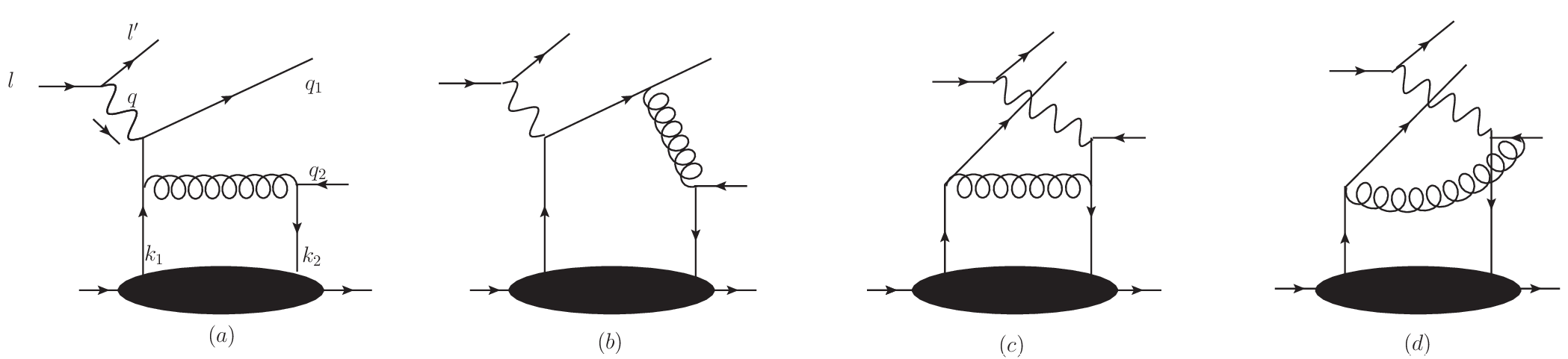}
\caption{Exclusive electroproduction of quark dijet at LO: dominant contribution from quark GPDs. }
\label{quark_dijetq}
\end{figure}
\begin{figure}[htbp]
\centering
\includegraphics[width=0.5\textwidth]{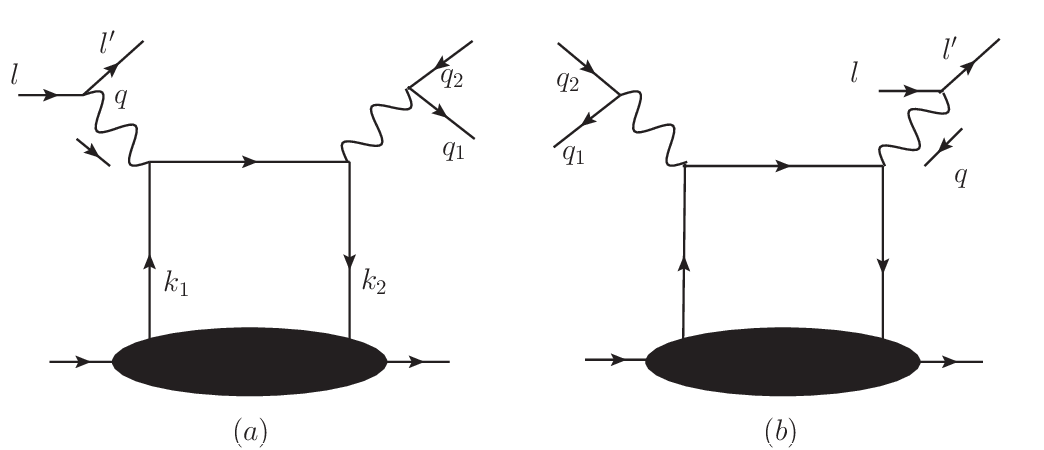}
\caption{Exclusive electroproduction of quark dijet at LO: subdominant contribution from quark GPDs. }
\label{quark_DDVCS}
\end{figure}

At LO, the amplitude receives contributions shown in Fig.\ref{quark_dijetq} and Fig.\ref{quark_DDVCS}. The quarks are taken to be massless in this section: $\mu^2=Q^2z\bar{z}.$ Note that the diagrams in Fig.~\ref{quark_DDVCS} are similar to the pure DDVCS subprocess~\cite{Deja:2023ahc}. In ~\cite{Deja:2023ahc}, a systematic study with EIC collision energy $10\text{GeV}\times100\text{GeV}$ was performed, and the contributions from the pure DDVCS subprocess were found to be much smaller than the Bethe-Heitler channel (which corresponds to the QED channel in quark dijet production). As will be shown in Sec.\ref{pheno}, the contributions from QED channel are subdominant, one can safely neglect the contributions from Fig.~\ref{quark_DDVCS}. The $k_1\,(k_2)$ denotes the momentum of the incoming (outgoing) parton. We have:
$$k_1^\mu=(x+\xi)Wn_+^\mu,\quad k_2^\mu=(x-\xi)Wn_+^\mu.$$

For contributions from unpolarized quark GPDs, we have:
\begin{align} \label{qqbaruq}
M^{q\bar{q},q,u}_{\text{QCD}}=&\frac{\sqrt{4\pi\alpha_{\text{em}}}}{Q^2}\cdot\frac{C\cdot C_F\xi W}{(\mathbf{q_\perp^2}+\mu^2)^2}\Big\{8Qz\bar{z}\bar{u}(l^\prime)\slashed{\epsilon}_Lu(l)\cdot\bar{u}(q_1)\slashed{n}_+v(q_2)I^{q\bar{q}}_{qu1}\nonumber \\
&+4\bar{u}(l^\prime)\gamma^\mu_\perp u(l)\Big[\bar{z}\bar{u}(q_1)\slashed{q}_\perp\gamma_{\perp\mu}\slashed{n}_+v(q_2)I_{qu2}^{q\bar{q}}+z\bar{u}(q_1)\gamma_{\perp\mu}\slashed{q}_\perp\slashed{n}_+v(q_2)I_{qu3}^{q\bar{q}}\Big]\Big\},
\end{align}
where $C=\big(2i\pi\alpha_s\sqrt{4\pi\alpha_{\text{em}}}e_q\big)/N_c$, $\gamma_\perp^\mu=g_{\perp}^{\mu\nu}\gamma_\nu.$ The Compton form factors (CFFs) $I_{qu1}^{q\bar{q}},I_{qu2}^{q\bar{q}},I_{qu3}^{q\bar{q}}$ are defined as:
\begin{align} \label{qqbaruCFF}
I^{q\bar{q}}_{qu1}=&\int_{-1}^1dxF_{qu}(x,\xi,t)\Big(\frac{z}{(x-\xi+i\epsilon)}+\frac{\bar{z}}{(x+\xi-i\epsilon)}\Big),\nonumber \\
I_{qu2}^{q\bar{q}}=&\int_{-1}^1dxF_{qu}(x,\xi,t)\Big(\frac{z}{x-\xi+i\epsilon}-\frac{\beta\bar{z}}{\bar{\beta}(x+\xi-i\epsilon)}+\frac{\bar{z}}{\bar{\beta}(x-\xi(1-2\beta)-i\epsilon)}\Big),\nonumber \\
I_{qu3}^{q\bar{q}}=&\int_{-1}^1dxF_{qu}(x,\xi,t)\Big(\frac{\beta z}{\bar{\beta}(x-\xi+i\epsilon)}-\frac{\bar{z}}{x+\xi-i\epsilon}-\frac{z}{\bar{\beta}(x+\xi(1-2\beta)+i\epsilon)}\Big).
\end{align}

One interesting feature in Eq.~\eqref{qqbaruCFF} is that when the photon is transversely polarized, the imaginary parts of CFFs receive contributions from the quark GPDs at $x\neq\xi$ (the last terms in $I_{qu2}^{q\bar{q}}$ and $I_{qu3}^{q\bar{q}}$).

The contributions from quark helicity GPDs can be obtained from Eq.~\eqref{qqbaruq} and Eq.~\eqref{qqbaruCFF} after the following replacements:
\begin{equation} \label{qqbarhq}
M^{q\bar{q},q,h}_{\text{QCD}}=M^{q\bar{q},q,u}_{\text{QCD}}\big(F_{qu}(x,\xi,t)\rightarrow F_{qh}(x,\xi,t),\slashed{n}_+\rightarrow \slashed{n}_+\gamma_5,\slashed{q}_\perp\slashed{e}_i\slashed{n}_+\rightarrow\slashed{q}_\perp\slashed{e}_i\slashed{n}_+\gamma_5,\slashed{e}_i\slashed{q}_\perp\slashed{n}_+\rightarrow\slashed{e}_i\slashed{q}_\perp\slashed{n}_+\gamma_5\big).
\end{equation}

\subsubsection{Contribution from gluon GPDs} \label{QCD_channel_ampg}
\begin{figure}[htbp]
\centering
\includegraphics[width=0.6\textwidth]{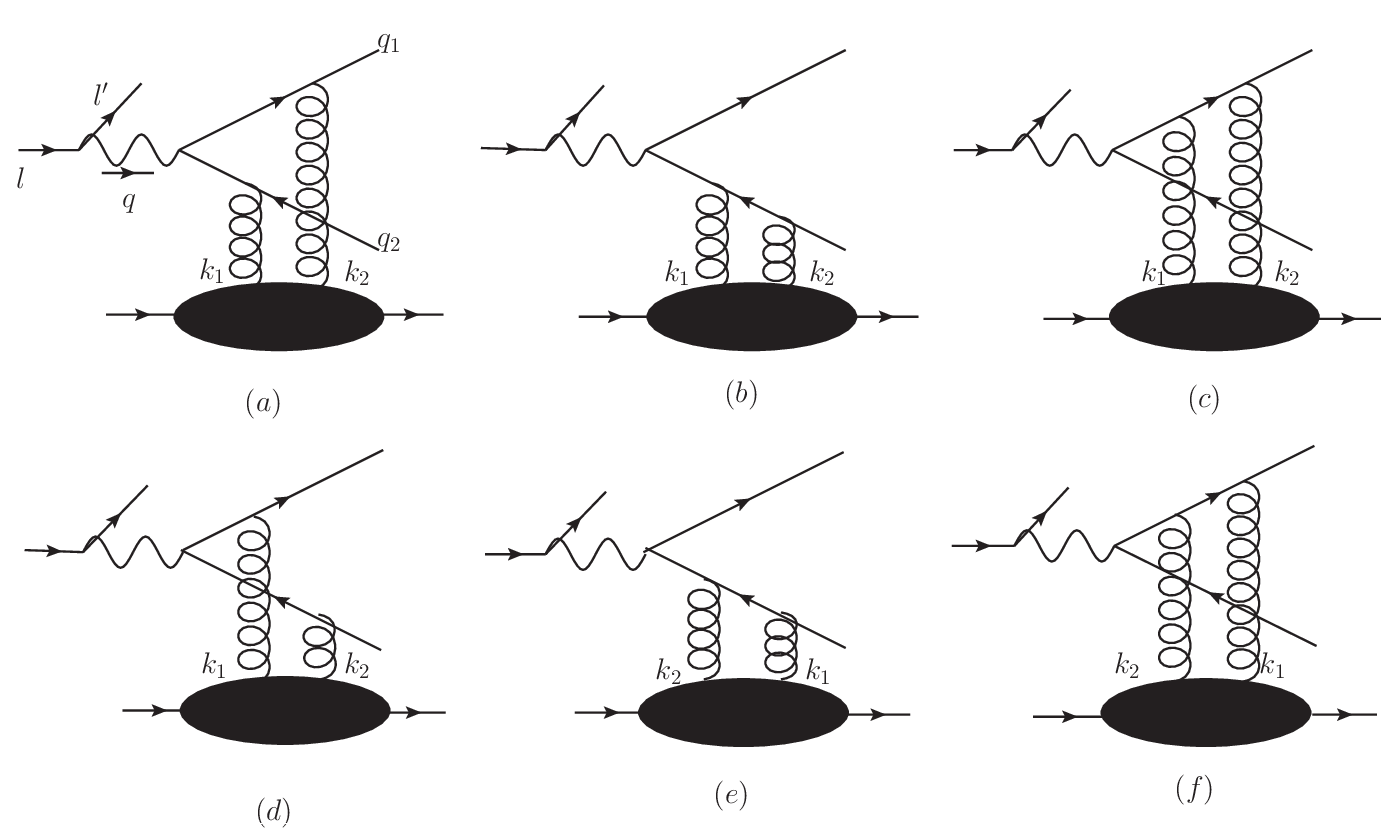}
\caption{Exclusive electroproduction of quark dijet at LO: contribution from gluon GPDs.}
\label{quark_dijetg}
\end{figure}
At LO, the amplitude receives contributions shown in Fig.\ref{quark_dijetg}. Note that due to Bose symmetry, there is a double-counting when adding the crossed diagrams, we need to multiply an extra $1/2$ factor in our calculations. We keep the quark mass: $\mu^2=Q^2\bar{z}z+m_q^2.$
For contributions from unpolarized gluon GPDs, we have:
\begin{align} \label{qqbarug}
M_{\text{QCD}}^{q\bar{q},g,u}=&\frac{\sqrt{4\pi\alpha_{\text{em}}}}{Q^2}\cdot\frac{C\xi W}{(\mu^2+\mathbf{q_\perp^2})^2}\Big\{4\bar{z}zQ\bar{u}(l^\prime)\slashed{\epsilon}_L u(l)\cdot\bar{u}(q_1)\slashed{n}_+v(q_2)I_{gu1}+\bar{u}(l^\prime)\gamma_\perp^\mu u(l)\Big[\big(\bar{z}\bar{u}(q_1)\slashed{q}_\perp\gamma_{\perp\mu}\slashed{n}_+v(q_2)\nonumber \\
&-z\bar{u}(q_1)\gamma_{\perp\mu}\slashed{q}_\perp\slashed{n}_+v(q_2)\big)I_{gu2}-2m_q\bar{u}(q_1)\gamma_{\perp\mu}\slashed{n}_+v(q_2)I_{gu1}\Big]\Big\},
\end{align}
where
\begin{align} \label{Igu12}
I_{gu1}=&\int_{-1}^1 dx F_{gu}(x,\xi,t)\Big(\frac{\bar{\beta}}{(x+\xi-i\epsilon)^2}+\frac{\bar{\beta}}{(x-\xi+i\epsilon)^2}-\frac{(1-2\beta)}{(x-\xi+i\epsilon)(x+\xi-i\epsilon)}\Big),\nonumber \\
I_{gu2}=&\int_{-1}^1 dxF_{gu}(x,\xi,t)\Big(\frac{(1-2\beta)}{(x+\xi-i\epsilon)^2}+\frac{(1-2\beta)}{(x-\xi+i\epsilon)^2}+\frac{4\beta}{(x-\xi+i\epsilon)(x+\xi-i\epsilon)}\Big).
\end{align}

For contributions from helicity gluon GPDs, we have:
\begin{align} \label{qqbarhg}
M_{\text{QCD}}^{q\bar{q},g,h}=&i\frac{\sqrt{4\pi\alpha_{\text{em}}}}{Q^2}\cdot\frac{2C\xi W\epsilon_{\perp\mu\nu}q_\perp^\nu}{(\mu^2+\mathbf{q_\perp^2})^2}\big(2\bar{z}\bar{u}(l^\prime)\gamma^\mu u(l)\cdot\bar{u}(q_1)\slashed{n}_+v(q_2)+\bar{u}(l^\prime)\gamma_\perp^\rho u(l)\cdot\bar{u}(q_1)\gamma_{\perp\rho}\slashed{n}_+\gamma^\mu v(q_2)\big)I_{gh},
\end{align}
where
\begin{equation} \label{Igh}
I_{gh}=\int_{-1}^1 dx F_{gh}(x,\xi,t)\Big(\frac{1}{(x-\xi+i\epsilon)^2}-\frac{1}{(x+\xi-i\epsilon)^2}\Big).
\end{equation}
Note that the contribution when the virtual photon is longitudinally polarized is zero in Eq.~\eqref{qqbarhg}. 

The gluon CFFs introduced above, ($I_{gu1},I_{gu2}$ and $I_{gh}$), contain terms with double poles at $x=\pm\xi$. After an integration by parts, such terms are proportional to the first derivative of gluon GPDs: $(\partial F_{gu/h}(x,\xi,t)/\partial x)/(x\pm\xi\mp i\epsilon)$. The first derivatives of gluon GPDs must be continuous at $x=\pm\xi$ such that the convolutions in Eq.~\eqref{Igu12} and Eq.~\eqref{Igh} are well-defined, 
which was found to be the case when the renormalization scale for gluon GPDs is
asymptotically large~\cite{Mankiewicz:1999tt,Radyushkin:1998es}. In a more general sense, such a condition was proved to be satisfied at an arbitrary renormalization scale using conformal partial wave expansion~~\cite{Mueller:2005ed}.
Consistent with these results, we have explicitly verified that the leading-order evolution of the first derivative of gluon GPDs doesn't generate discontinuities at $x=\pm\xi$, the details can be found in Appendix~A.

It is worth pointing out that for the QCD channel, the amplitudes for transversely polarized photon and longitudinally polarized photon show different behaviors at small $\big|q_\perp\big|$. When the photon is transversely (longitudinally) polarized, the amplitude vanishes (is finite) in the limit $\big|q_\perp\big|\rightarrow0$, for production of light flavored dijet ($f=u,d,s$). This observation can be explained by the conservation of total angular momentum along $z$ direction.

\subsubsection{Contribution from elastic FFs} \label{QED_amp}
\begin{figure}[htbp]
\centering
\includegraphics[width=1.0\textwidth]{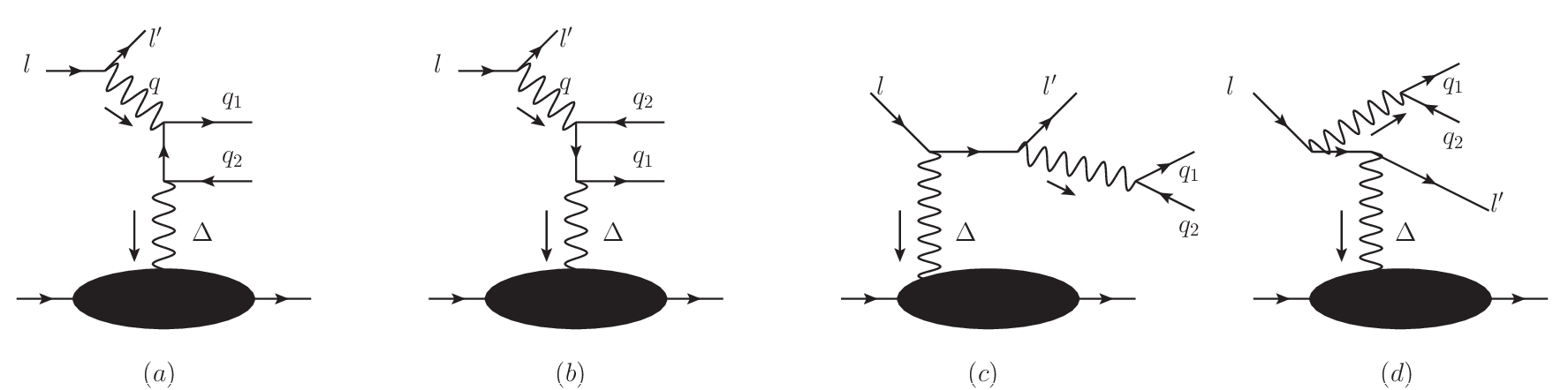}
\caption{Exclusive electroproduction of quark dijet at LO: contribution from EFFs.}
\label{primakofffig}
\end{figure}
The amplitude for this channel can be written as:
\begin{equation} \label{primakoff}
M_{\text{QED}}^{q\bar{q}}=\Sigma_fe_fH^\alpha \frac{g_{\alpha\beta}}{t+i\epsilon}J_f^\beta,
\end{equation}
where $H^{\alpha}$ denotes the hard scattering part, $e_f$ is the charge of the quark with flavor $f$, $J^\alpha_f$ is defined by Eq.~\eqref{EFF}.

To obtain the leading-power contribution to Eq.~\eqref{primakoff}, we use the following Grammer-Yennie decomposition ~\cite{Grammer:1973db}:
\begin{equation}
g_{\alpha\beta}=G_{\alpha\beta}+K_{\alpha\beta},\quad G_{\alpha\beta}=g_{\alpha\beta}-\frac{\Delta_\alpha n_{-\beta}}{\Delta\cdot n_-},\quad K_{\alpha\beta}=\frac{\Delta_\alpha n_{-\beta}}{\Delta\cdot n_-},
\end{equation}
which allows us to rewrite Eq.~\eqref{primakoff} as:
\begin{equation} \label{primakoffG}
M_{\text{QED}}^{q\bar{q}}=\Sigma_fe_fH^\alpha \frac{G_{\alpha\beta}}{t+i\epsilon}J_f^\beta,
\end{equation}
where Ward identity has been used. The leading-power contribution to Eq.~\eqref{primakoffG} can be obtained by taking $\alpha$ to be transverse. In the following calculation of amplitude, we use the shorthand notation $\tilde{J}_\alpha=G_{\alpha\beta}J^\beta.$

The amplitude for Fig.~\ref{primakofffig} reads:
\begin{align} \label{Primakoff}
M_{\text{QED}}^{q\bar{q}}=&\Sigma_fe_f\frac{i(4\pi\alpha_{\text{em}})^2e_q\tilde{J}_{f,\alpha}}{t(\mu^2+\mathbf{q_\perp^2})^2}\Big\{\frac{e_q}{Q^2}\Big[2\xi Q^2W\bar{z}z^2\bar{u}(q_1)\gamma^\mu \slashed{n_+}\gamma^\alpha v(q_2)-\frac{(\mu^2+\mathbf{q_\perp^2})^2}{4\xi W\bar{z}}\bar{u}(q_1)\gamma^\mu \slashed{n_-}\gamma^\alpha v(q_2)\nonumber \\
&-2\xi W\bar{z}(\mu^2+\mathbf{q_\perp^2})\bar{u}(q_1)\gamma^\alpha \slashed{n_+}\gamma^\mu v(q_2)+2z(\mu^2+\mathbf{q_\perp^2})q_1^\mu\bar{u}(q_1)\gamma^\alpha v(q_2)+2\bar{z}(\mu^2+\mathbf{q_\perp^2})q_\perp^\alpha\bar{u}(q_1)\gamma^\mu v(q_2)\Big]\nonumber \\
&\cdot\bar{u}(l^\prime)\gamma_\mu u(l)+\frac{2(\mu^2+\mathbf{q_\perp^2})y\bar{z}^2z^2}{(1-y)(m_q^2+\mathbf{q_\perp^2})}\Big[-\xi W\bar{u}(l^\prime)\gamma^\alpha\slashed{n_+}\gamma^\mu u(l)-\xi W(1-y)\bar{u}(l^\prime)\gamma^\mu\slashed{n_+}\gamma^\alpha u(l)+y l_\perp^\alpha \bar{u}(l^\prime)\gamma^\mu u(l)\Big]\nonumber \\
&\cdot\bar{u}(q_1)\gamma_\mu v(q_2)\Big\}.
\end{align}
We keep the nonzero value of $m_q$ in Eq.~\eqref{Primakoff} only for the production of charm and bottom quark dijets. Unlike the Bethe-Heitler channel in DDVCS, the charge structures in Fig.(a)/Fig.(b) and Fig.(c)/Fig.(d) are different. Due to contributions from the last two diagrams in Fig.~\ref{primakofffig}, there is a collinear divergence in $M_{\text{QED}}^{q\bar{q}}$ as $\big|q_\perp\big|\rightarrow0$ for the production of light flavored dijet.

\subsection{Gluon dijet production}
\begin{figure}[htbp]
\centering
\includegraphics[width=0.6\textwidth]{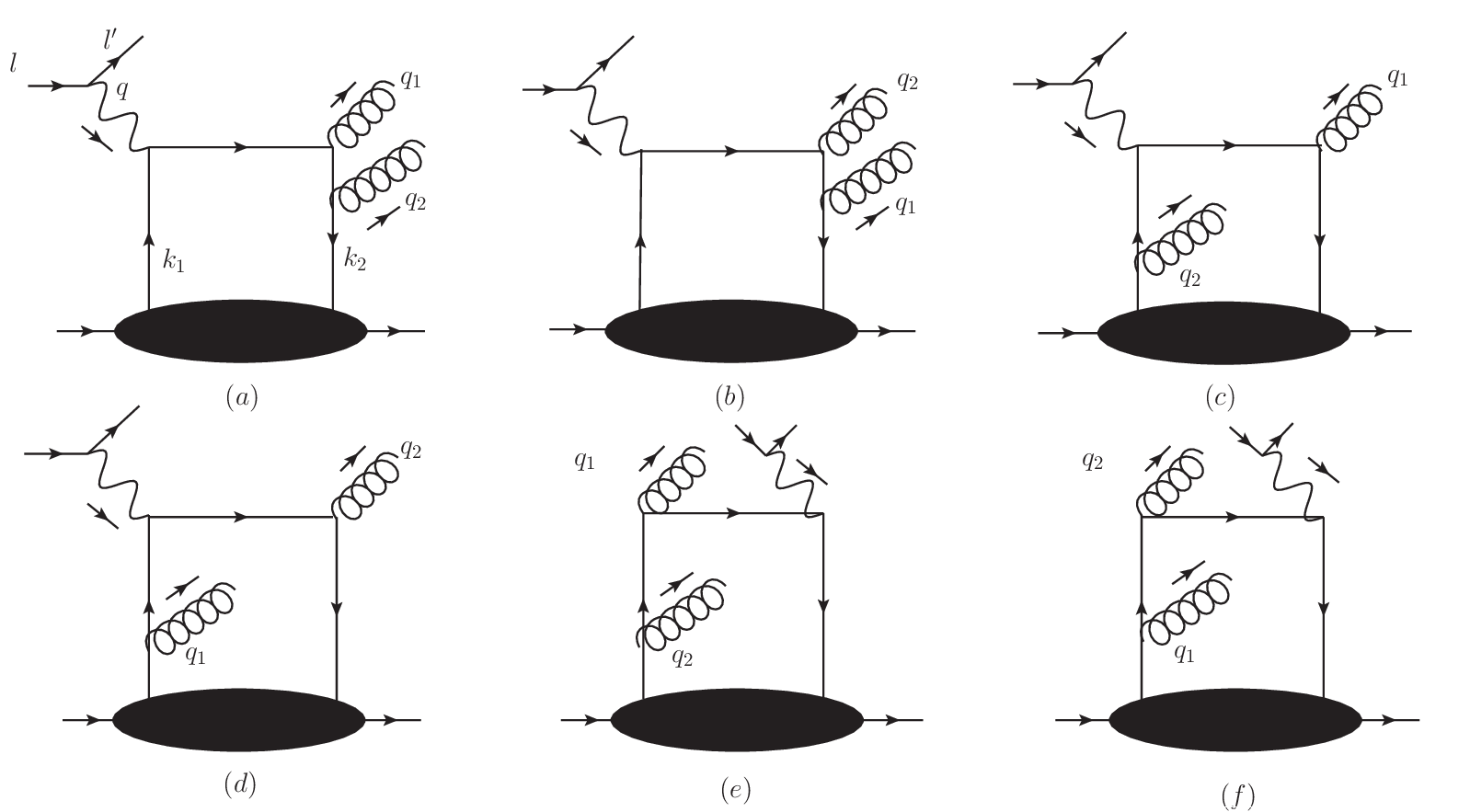}
\caption{Exclusive electroproduction of gluon dijet at LO.}
\label{gluon_dijet}
\end{figure}

At LO in $\alpha_s$, the contributions are shown in Fig.\ref{gluon_dijet}.
Note that due to charge conjugation symmetry, the exclusive gluon dijet production is only sensitive to quark C-odd GPDs (in other words, quark valence GPDs) at any order in perturbation theory~\cite{Pedrak:2017cpp,Pedrak:2020mfm}. To simplify the calculations in this section, we work in the gauge $n_+\cdot A=0$, where the two gluon polarization vectors $\epsilon_{1,i}$ and $\epsilon_{2,i}$ take the following form~\cite{Grocholski:2021man}(the subscript $``i"$ denotes the helicity):
\begin{align}
\epsilon_{1,i}^{\mu}=&\epsilon^\mu_{1\perp,i}-2\frac{q_\perp\cdot\epsilon_{1\perp,i}}{z(Q^2+W^2)}W(1+\xi)n_+^\mu,\nonumber \\
\epsilon_{2,i}^\mu=&\epsilon^\mu_{2\perp,i}+2\frac{q_\perp\cdot\epsilon_{2\perp,i}}{\bar{z}(Q^2+W^2)}W(1+\xi)n_+^\mu,
\end{align}
For contribution from unpolarized quark GPDs, we have:
\begin{align} \label{ggu}
M^{gg,u}=&\Sigma_f\frac{\sqrt{4\pi\alpha_{\text{em}}}}{Q^2}C_f\cdot C_F\Big\{\bar{u}(l^\prime)\slashed{\epsilon}_Lu(l)\cdot\frac{8 Q \bar{z} z (\epsilon_{1\perp,i}\cdot\epsilon_{2\perp,j})}{\left(\mu^2+\mathbf{q_\perp^2}\right)}I^{gg}_{qu1,f}\nonumber \\
&+\bar{u}(l^\prime)\gamma_{\perp\mu} u(l)\cdot\frac{2\big((2z-1)q_\perp^\mu(\epsilon_{1\perp,i}\cdot\epsilon_{2\perp,j})+(q_\perp\cdot\epsilon_{2\perp,j})\epsilon_{1\perp,i}^\mu-(q_\perp\cdot\epsilon_{1\perp,i})\epsilon_{2\perp,j}^\mu\big)}{\mathbf{q_\perp^2}}I_{qu2,f}^{gg}\Big\},
\end{align}
where the colors of the produced gluon pair and the flavors ($f=u,d,s$) of the active quarks have been summed over, we have $C_f=2i\pi\alpha_s\sqrt{4\pi\alpha_{\text{em}}}e_f$ with $e_f$ the charge of quark with flavor $f.$ 
The CFFs in Eq.~\eqref{ggu} are defined as:
\begin{align} \label{Iqu1f}
I^{gg}_{qu1,f}=&\int_{-1}^1dxF_{qu,f}(x,\xi,t)\Big(\frac{1}{x+\xi-i\epsilon}-\frac{1}{x-\xi+i\epsilon}\Big),\nonumber \\
I^{gg}_{qu2,f}=&\int_{-1}^1dxF_{qu,f}(x,\xi,t)\Big((1-2\beta)\big(\frac{1}{x-\xi+i\epsilon}-\frac{1}{x+\xi-i\epsilon}\big)+\frac{1}{x+\xi(1-2\beta)+i\epsilon}-\frac{1}{x-\xi(1-2\beta)-i\epsilon}\Big),
\end{align}
where $F_{qu,f}$ denotes the $f-$flavored quark unpolarized GPD. 

For contributions from helicity quark GPDs, 
we use the following variation of Schouten identity:
\begin{equation}
\epsilon_{1\perp,i}\cdot\epsilon_{2\perp,j}\epsilon^{n_-n_+q_\perp e_i}=\epsilon_{1\perp,i}\cdot e_i\epsilon^{n_-n_+q_\perp \epsilon_{2\perp,j}}-\epsilon_{1\perp,i}\cdot q_\perp\epsilon^{n_-n_+e_i\epsilon_{2\perp,j}},
\end{equation}
which holds for $e_i=e_{x,y}$ and general $q_\perp,\epsilon_{1\perp,i},\epsilon_{2\perp,j}$. The results can be written as:
\begin{equation} \label{ggh1}
M^{gg,h}=-\Sigma_f\frac{2i\sqrt{4\pi\alpha_{\text{em}}}C_f\cdot C_F}{Q^2\mathbf{q_\perp^2}}\bar{u}(l^\prime)\gamma_{\perp\mu} u(l)\Big[q_\perp^\mu\epsilon^{n_-n_+\epsilon_{1\perp,i}\epsilon_{2\perp,j}}-(2z-1)(\epsilon_{1\perp,i}\cdot\epsilon_{2\perp,j})\epsilon^{\mu n_-n_+q_\perp }\Big]I_{qh1,f}^{gg},
\end{equation}
the CFF $I_{qh1,f}^{gg}$ is defined as:
\begin{equation} \label{Iqh1f}
I_{qh1,f}^{gg}=\int_{-1}^1dxF_{qh,f}(x,\xi,t)\Big[\frac{1}{x+\xi-i\epsilon}+\frac{1}{x-\xi+i\epsilon}-\frac{1}{x-\xi(1-2\beta)-i\epsilon}-\frac{1}{x+\xi(1-2\beta)+i\epsilon}\Big].
\end{equation}
Two comments on the above results are in order. 
First of all, the hard parts in Eq.~\eqref{Iqu1f} and Eq.~\eqref{Iqh1f} are odd in $(x\rightarrow -x)$ for contributions from unpolarized GPDs and even in $(x\rightarrow -x)$ for contributions from helicity GPDs, so only C-odd quark GPDs contribute to the gluon dijet production, which is consistent with the expectation from charge conjugation symmetry. Secondly, similar to quark dijet production, the CFFs in Eq.~\eqref{Iqu1f} and Eq.~\eqref{Iqh1f} are sensitive to valence quark GPDs at $x\neq\xi$ (the sensitivity originates from the last two terms in $I_{qu2,f}^{gg}$ and $I_{qh1,f}^{gg}$). This is due to the fact that for the electroproduction of dijets, there are multiple hard scales that enter the calculation: $Q^2$, $\mathbf{q_\perp^2}$, and $m_q^2$ (nonzero for the production of charm and bottom dijets).

\section{Differential cross sections and other physical observables}
\label{diff cross section}

In this paper, we consider the case when both beam and target are unpolarized. The fully differential cross section can be written as \cite{Braun:2005rg}:
\begin{align} \label{master}
\frac{d^8\sigma}{dydzdQ^2d\mathbf{q_\perp^2}d\mathbf{\Delta_\perp^2}d\phi d\phi_{l^\prime}d\phi_{p^\prime}}=&\frac{y}{2^{15}\pi^8\bar{z}z(W^2-M^2)(Q^2+W^2)}\cdot\frac{1}{2}\Sigma_{\lambda^\prime,\lambda}\big|M\big|^2,
\end{align}
where the helicity of the initial($\lambda$) (final($\lambda^{\prime}$)) proton is averaged (summed over).

\subsection{Quark dijet production}
For quark dijet production, the $\big|M^{q\bar{q}}\big|^2$ in Eq.~\eqref{master} can be written as a sum of three terms:
\begin{equation} \label{sumQEDQCD}
\big|M^{q\bar{q}}\big|^2=\big|M^{q\bar{q}}_{\text{QCD}}\big|^2+2\text{Re}\big(M^{q\bar{q}}_{\text{QCD}}M^{q\bar{q}*}_{\text{QED}}\big)+\big|M^{q\bar{q}}_{\text{QED}}\big|^2,
\end{equation}
which represent contributions from the pure QCD channel, interference between the QCD and QED channel, and the pure QED channel, respectively. For brevity, we refer to them as QCD, interference, and QED contributions in the following sections.
The expressions for $2\text{Re}(M_{QCD}^{q\bar{q}}M_{QED}^{q\bar{q}*})$ and $\big|M_{QED}^{q\bar{q}}\big|^2$ are given in Appendix B.

For unpolarized target, $\big|M^{q\bar{q}}_{\text{QCD}}\big|^2$ can be written as a sum of two terms:
\begin{equation}
\big|M^{q\bar{q}}_{\text{QCD}}\big|^2=\big|M^{q\bar{q},uu}_{\text{QCD}}\big|^2+\big|M^{q\bar{q},hh}_{\text{QCD}}\big|^2,
\end{equation}
where $``\big|M^{q\bar{q},aa}_{\text{QCD}}\big|^2"$ denotes the contribution $\big(M^{q\bar{q},q,a}_{\text{QCD}}+M^{q\bar{q},g,a}_{\text{QCD}}\big)\big(M^{q\bar{q},q,a*}_{\text{QCD}}+M^{q\bar{q},g,a*}_{\text{QCD}}\big)$ with $aa\in\big\{uu,hh\big\}.$ The specific expressions of $M^{q\bar{q},q,a}_{\text{QCD}}$ and $M^{q\bar{q},g,a}_{\text{QCD}}$ are given by Eq.~\eqref{qqbaruq}, Eq.~\eqref{qqbarhq}, Eq.~\eqref{qqbarug} and Eq.~\eqref{qqbarhg}.

$\big|M^{q\bar{q},aa}_{\text{QCD}}\big|^2$ has the following structure:
\begin{equation} \label{qqbarQCD}
\big|M^{q\bar{q},aa}_{\text{QCD}}\big|^2=\frac{2\pi\alpha_{\text{em}}}{Q^4}\Big[\frac{8Q^2(1-y)}{y^2}\big|\mathcal{A}^{q\bar{q}}_L\big|^2+\frac{8Q^2\sqrt{1-y}(2-y)}{y^2}\text{Re}\big(\mathcal{A}_T^{q\bar{q},x}\mathcal{A}_L^{q\bar{q}*}\big)+\frac{2Q^2(2-y)^2}{y^2}\big|\mathcal{A}_T^{q\bar{q},x}\big|^2+2Q^2\big|\mathcal{A}_T^{q\bar{q},y}\big|^2\Big],
\end{equation}

For $\big|M^{q\bar{q},uu}_{\text{QCD}}\big|^2$ we have:
\begin{align} \label{csQCDuu}
\big|\mathcal{A}_T^{q\bar{q},x}\big|^2=&\frac{\big|C\big|^2N_c}{(\mu^2+\mathbf{q_\perp^2})^2}\Big[\frac{2m_q^2}{\bar{z}z}\big|I_{gu1}\big|^2+\frac{\bar{z}\mathbf{q_\perp^2}}{2z}\big|4C_FI_{qu2}^{q\bar{q}}+I_{gu2}\big|^2+\frac{z\mathbf{q_\perp^2}}{2\bar{z}}\big|4C_FI_{qu3}^{q\bar{q}}-I_{gu2}\big|^2\nonumber \\
&+\text{cos}(2\phi)\mathbf{q_\perp^2}\text{Re}\big((4C_FI_{qu2}^{q\bar{q}}+I_{gu2})(4C_FI_{qu3}^{q\bar{q}}-I_{gu2})^*\big)\Big],\nonumber \\
\big|\mathcal{A}_T^{q\bar{q},y}\big|^2=&\big|\mathcal{A}_T^{q\bar{q},x}\big|^2\big(\text{cos}(2\phi)\rightarrow-\text{cos}(2\phi)\big),\nonumber \\
\text{Re}(\mathcal{A}_T^{q\bar{q},x}\mathcal{A}_L^{q\bar{q}*})=&-\frac{2\big|C\big|^2N_c\text{cos}(\phi)Q\big|q_\perp\big|}{(\mu^2+\mathbf{q_\perp^2})^2}\Big[\bar{z}\text{Re}\big((4C_FI_{qu2}^{q\bar{q}}+I_{gu2})(2C_FI_{qu1}^{q\bar{q}}+I_{gu1})^*\big)\nonumber \\
&+z\text{Re}\big((4C_FI_{qu3}^{q\bar{q}}-I_{gu2})(2C_FI_{qu1}^{q\bar{q}}+I_{gu1})^*\big)\Big],\nonumber \\
\big|\mathcal{A}_L^{q\bar{q}}\big|^2=&\frac{8\big|C\big|^2N_cQ^2\bar{z}z}{(\mu^2+\mathbf{q_\perp^2})^2}\big|2C_FI_{qu1}^{q\bar{q}}+I_{gu1}\big|^2.
\end{align}

For $\big|M^{q\bar{q},hh}_{\text{QCD}}\big|^2$ we have ($I_{qhi}^{q\bar{q}}=I_{qui}^{q\bar{q}}(F_{qu}(x,\xi,t)\rightarrow F_{qh}(x,\xi,t)),i=1,2,3$):
\begin{align} \label{csQCDhh}
\big|\mathcal{A}_T^{q\bar{q},x}\big|^2=&\frac{8\big|C\big|^2N_c\mathbf{q_\perp^2}}{(\mu^2+\mathbf{q_\perp^2})^2}\Big[\frac{C_F^2\bar{z}}{z}\big|I^{q\bar{q}}_{qh2}\big|^2+\frac{C_F^2z}{\bar{z}}\big|I^{q\bar{q}}_{qh3}\big|^2+\frac{\big(2\bar{z}z(\text{cos}(2\phi)-1)+1\big)}{4\bar{z}z}\big|I_{gh}\big|^2+2C_F^2\text{cos}(2\phi)\text{Re}\big(I^{q\bar{q}}_{qh2}I^{q\bar{q}*}_{qh3}\big)\nonumber \\
&-\frac{C_F\big(z(1-\text{cos}(2\phi))-1\big)}{z}\text{Re}\big(I^{q\bar{q}}_{qh2}I_{gh}^*\big)-\frac{C_F\big(\text{cos}(2\phi)(z-1)-z\big)}{\bar{z}}\text{Re}\big(I^{q\bar{q}}_{qh3}I_{gh}^*\big)\Big],\nonumber \\
\big|\mathcal{A}_T^{q\bar{q},y}\big|^2=&\big|\mathcal{A}_T^{q\bar{q},x}\big|^2\big(\text{cos}(2\phi)\rightarrow-\text{cos}(2\phi)\big),\nonumber \\
\text{Re}\big(\mathcal{A}_T^{q\bar{q},x}\mathcal{A}^{q\bar{q}*}_L\big)=&\frac{16\big|C\big|^2N_c}{(\mu^2+\mathbf{q_\perp^2})^2}\Big[-C_F^2Q\big|q_\perp\big|\bar{z}\text{cos}(\phi)\text{Re}(I^{q\bar{q}}_{qh1}I^{q\bar{q}*}_{qh2})-C_F^2Q\big|q_\perp\big|z\text{cos}(\phi)\text{Re}(I^{q\bar{q}}_{qh1}I^{q\bar{q}*}_{qh3})\nonumber \\&-\frac{1}{2}C_FQ\big|q_\perp\big|\text{cos}(\phi)\text{Re}(I^{q\bar{q}}_{qh1}I_{gh}^*)\Big],\nonumber \\
\big|\mathcal{A}_L^{q\bar{q}}\big|^2=&\frac{32\big|C\big|^2N_c\bar{z}z}{(\mu^2+\mathbf{q_\perp^2})^2}C_F^2Q^2\big|I^{q\bar{q}}_{qh1}\big|^2.
\end{align}
The $m_q$'s in Eq.~\eqref{csQCDuu} and Eq.~\eqref{csQCDhh} are nonzero only when the two conditions are both satisfied: (1).The $m_q$'s appear in the coefficient of interference between a quark CFF and a quark CFF or gluon CFF.
(2). When it is the charm or bottom quark dijet that is produced.

\subsection{Gluon dijet production}
For unpolarized target, similar to Eq.~\eqref{qqbarQCD}, $\big|M^{gg}\big|^2$ can be written as a sum of two terms:
\begin{align}
\big|M^{gg}\big|^2=\big|M^{gg,uu}\big|^2+\big|M^{gg,hh}\big|^2,
\end{align}
where $\big|M^{gg,aa}\big|^2$ has the following structure:
\begin{equation}
\big|M^{gg,aa}\big|^2=\frac{2\pi\alpha_{\text{em}}}{Q^4}\Big[\frac{8Q^2(1-y)}{y^2}\big|\mathcal{A}^{gg}_L\big|^2+\frac{8Q^2\sqrt{1-y}(2-y)}{y^2}\text{Re}\big(\mathcal{A}_T^{gg,x}\mathcal{A}_L^{gg*}\big)+\frac{2Q^2(2-y)^2}{y^2}\big|\mathcal{A}_T^{gg,x}\big|^2+2Q^2\big|\mathcal{A}_T^{gg,y}\big|^2\Big].
\end{equation}

For $\big|M^{gg,uu}\big|^2$ we have:
\begin{align} \label{gguu}
\big|\mathcal{A}_T^{gg,x}\big|^2=&\frac{8C_F^2}{\mathbf{q_\perp^2}}\Big[-2\text{cos}(2\phi)z\bar{z}-2\bar{z}z+1\Big](\Sigma_fC_fI_{qu2,f}^{gg})(\Sigma_{f^\prime}C_{f^\prime}^*I_{qu2,f^{\prime}}^{gg*}),\nonumber \\
\big|\mathcal{A}_T^{gg,y}\big|^2=&\big|\mathcal{A}_T^{gg,x}\big|^2\big(\text{cos}(2\phi)\rightarrow-\text{cos}(2\phi)\big),\nonumber \\
\text{Re}\big(\mathcal{A}_T^{gg,x}\mathcal{A}_L^{gg*}\big)=&\frac{32C_F^2Qz\bar{z}(1-2z)\text{cos}(\phi)}{\big|q_\perp\big|(\mu^2+\mathbf{q_\perp^2})}\text{Re}\Big[(\Sigma_fC_fI_{qu1,f}^{gg})(\Sigma_{f^\prime}C^*_{f^\prime}I_{qu2,f^\prime}^{gg*})\Big],\nonumber \\
\big|\mathcal{A}_L^{gg}\big|^2=&\frac{128C_F^2Q^2\bar{z}^2z^2}{(\mu^2+\mathbf{q_\perp^2})^2}(\Sigma_fC_fI_{qu1,f}^{gg})(\Sigma_{f^\prime}C_{f^\prime}^*I_{qu1,f^{\prime}}^{gg*}).
\end{align}

For $\big|M^{gg,hh}\big|^2$ we have:
\begin{align} \label{gghh}
\big|\mathcal{A}_T^{gg,x}\big|^2=&\frac{8C_F^2}{\mathbf{q_\perp^2}}\Big[2\text{cos}(2\phi)z\bar{z}-2z\bar{z}+1\Big](\Sigma_fC_fI_{qh1,f}^{gg})(\Sigma_{f^\prime}C_{f^\prime}^*I_{qh1,f^{\prime}}^{gg*}),\nonumber \\
\big|\mathcal{A}_T^{gg,y}\big|^2=&\big|\mathcal{A}_T^{gg,x}\big|^2\big(\text{cos}(2\phi)\rightarrow-\text{cos}(2\phi)\big),\nonumber \\
\text{Re}\big(\mathcal{A}_T^{gg,x}\mathcal{A}_L^{gg*}\big)=&0,\nonumber \\
\big|\mathcal{A}_L^{gg}\big|^2=&0.
\end{align}
To derive Eq.~\eqref{gguu} and Eq.~\eqref{gghh}, we have used $\Sigma_i\epsilon_{1(2)\perp,i}^\mu\epsilon_{1(2)\perp,i}^{*\nu}=-g^{\mu\nu}_\perp.$ We have $\mu^2=Q^2z\bar{z}$ in Eq.~\eqref{gguu} and Eq.~\eqref{gghh}.
\section{Numerical Results}
\label{pheno}
\begin{figure}[htbp]
\centering
\includegraphics[width=1.0\textwidth]{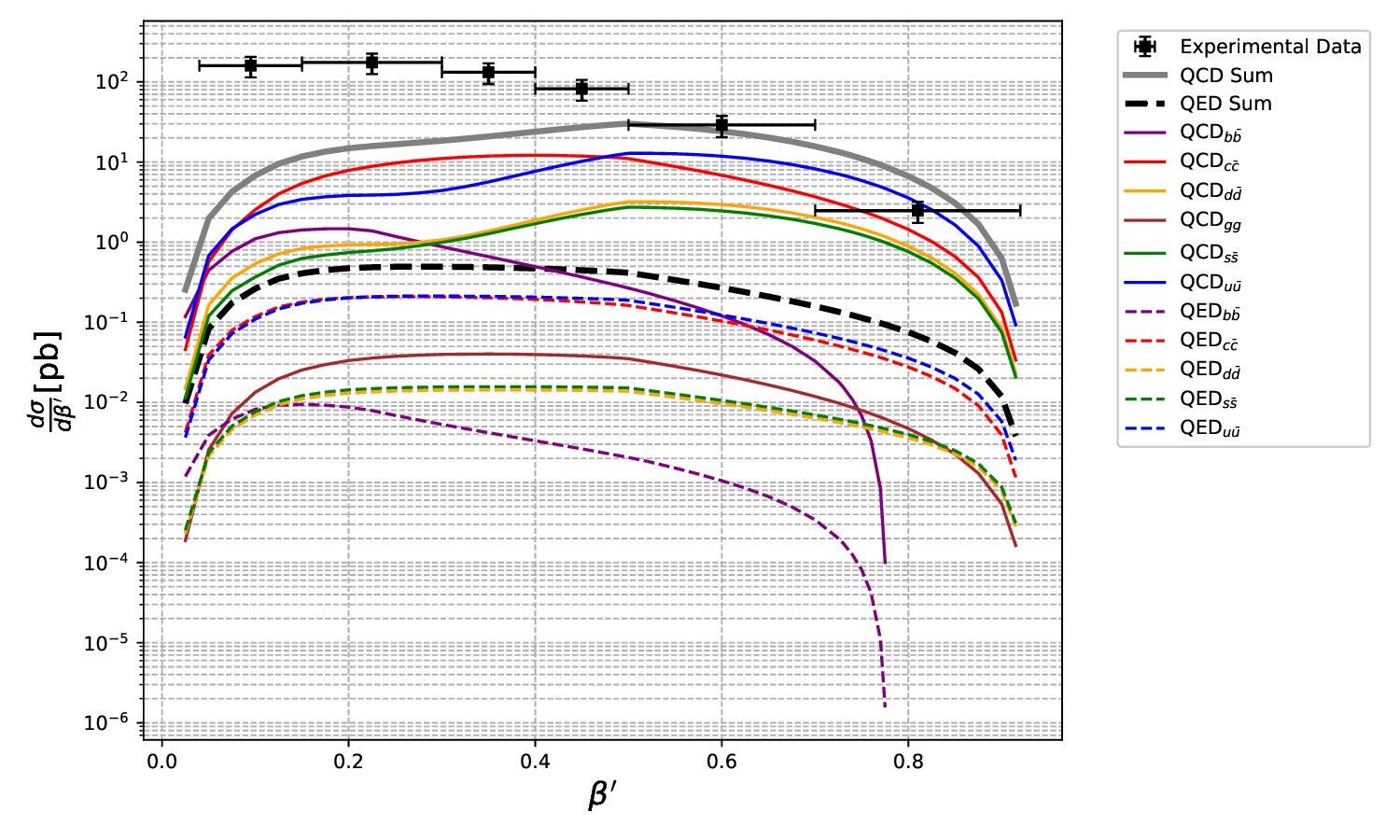}
\caption{The cross section $d\sigma/d\beta^\prime$ as a function of $\beta^\prime$ in HERA kinematics. The six black bins denote the data from the ZEUS collaboration with the center-of-mass energy $\sqrt{s}=318\text{GeV}$~\cite{ZEUS:2015sns}. The solid and dashed lines denote the QCD and QED contributions, respectively. The subscripts denote the types of dijets that are produced: quark dijets or gluon dijets.
The contributions from the production of different dijets are shown separately. The following kinematic cuts are adopted in the numerical integration~\cite{ZEUS:2015sns}: $0.1<y<0.64$, $-1.5\,\text{GeV}^2<t<-0.01\,\text{GeV}^2$, $4\,\text{GeV}^2<\mathbf{q_\perp^2}<50\,\text{GeV}^2$, $25\,\text{GeV}^2<Q^2<300\,\text{GeV}^2$, $90\,\text{GeV}<W<250\,\text{GeV}$, $0<x_{\mathbb{P}}(\equiv x_{\text{Bj}}/\beta^\prime)<0.01$, $M>5\,\text{GeV}$, $\eta<2$, the integrations over $\phi$ and $\phi_{p^\prime}$ are performed over the full range. The definition for pseudorapidity $\eta$ can be found in ~\cite{ZEUS:2015sns}. The sum of QED contributions and QCD contributions is not shown, since the former is much smaller than the latter.}
\label{beta_HERA}
\end{figure}
\begin{figure}[htbp]
\centering
\includegraphics[width=1.0\textwidth]{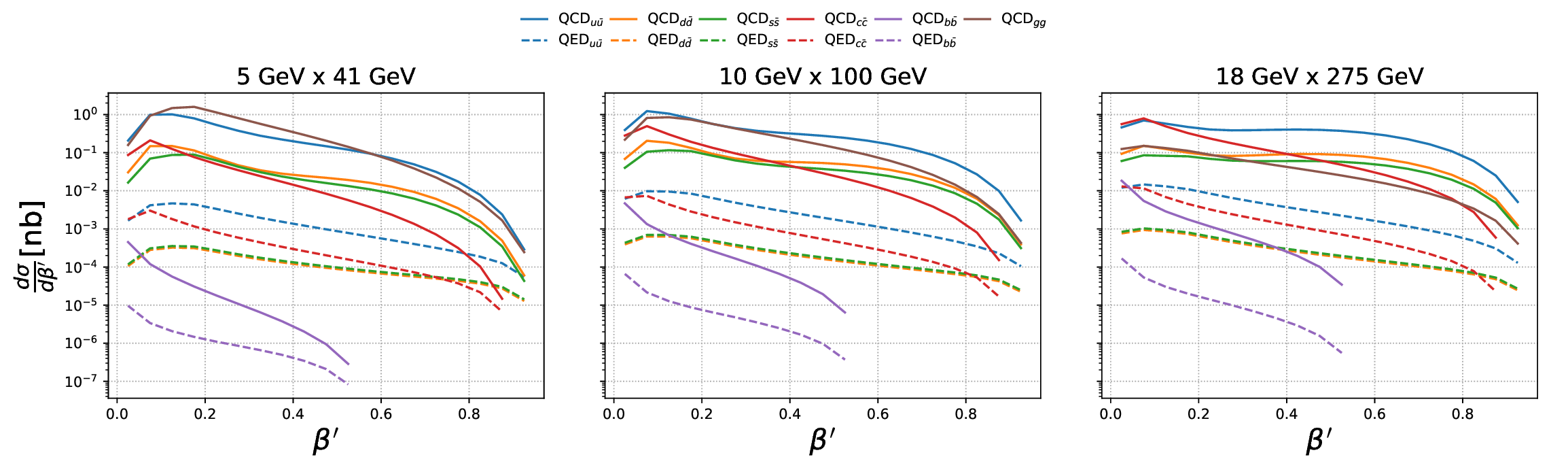}
\caption{The cross section $d\sigma/d\beta^\prime$ as a function of $\beta^\prime$ for different EIC collision energies. The following kinematic cuts are used in the numerical integration: $0.01<y<0.95$, $-1.5\,\text{GeV}^2<t<-0.01\,\text{GeV}^2$, $1\,\text{GeV}^2<\mathbf{q_\perp^2}<50\,\text{GeV}^2$, $0.05<z<0.95$, $1\,\text{GeV}^2<Q^2<100\,\text{GeV}^2$. The integrations over $\phi$ and $\phi_{p^\prime}$ are performed over the full range.}
\label{beta_EIC}
\end{figure}

Having established the theoretical framework, we now turn to its phenomenological implications. In this section, we present predictions for quark- and gluon-dijet production in HERA and EIC kinematics, comparing the former with available HERA data and assessing the prospects for measurements at the future EIC. The numerical results are obtained using the Goloskokov--Kroll GPD model~\cite{Goloskokov:2006hr,Goloskokov:2007nt}. In the context of $k_t$ factorization of quark dijet production, the typical scale defining the hardness of the Pomeron exchanged between the virtual photon and the proton was found to be $m_q^2+\mathbf{q_\perp^2}+z\bar{z}Q^2$~\cite{Bartels:1996ne}. In our analyses below, we adopt this scale to be
the renormalization and factorization scales: $\mu_{\text{R}}^2=\mu_{\text{F}}^2=m_q^2+\mathbf{q_\perp^2}+z\bar{z}Q^2$. For simplicity, we only consider contributions from those GPDs which are dominant: $H^{q,g}$ (for unpolarized GPD) and $\tilde{H}^{q,g}$ (for helicity GPD). In comparison, the magnitude of $E^{q,g}$($\tilde{E}^{q,g}$) is much smaller than $H^{q,g}$($\tilde{H}^{q,g}$) and will be neglected.

Figures~\ref{beta_HERA} and~\ref{beta_EIC} show the differential cross section as a function of $\beta^\prime$ (defined by Eq.~\eqref{betadiff}) in HERA and EIC kinematics, respectively. Note that for the QED contributions, the production of down (shown in orange dashed lines) and strange quark (shown in green dashed lines) dijets gives the same cross section, since they have the same charge (see Sec.~\ref{inter_form}). For visual clarity, the green dashed lines are slightly shifted in Fig.~\ref{beta_HERA} and Fig.~\ref{beta_EIC}.
In HERA kinematics, our theoretical prediction is consistent with the data at $\beta^\prime\gtrsim 0.5$; however, it fails to describe the data
at $\beta^\prime\lesssim 0.5$. This behavior can be understood from the definition of $\beta^\prime$ (Eq.~\eqref{betadiff}): at smaller values of $\beta^\prime$, the invariant mass ($M$) of the dijet system tends to get larger, enhancing the phase space available for real gluon radiation~\cite{Bartels:1999tn,Boussarie:2019ero}. 
In HERA kinematics, the contribution of gluon dijet production (shown in the solid brown line) remains significantly smaller than the contribution of quark dijet production over the entire $\beta^\prime$ region. In contrast, in EIC kinematics (shown in Fig.~\ref{beta_EIC}) with colliding energy $5\,\text{GeV}\times 41\,\text{GeV}$, the contribution of gluon dijet production is comparable to the contribution of up quark dijet production (shown in the solid blue lines). This difference can be attributed to the different collision energies and coverage of $y$: in the region of smaller $y$ ($0.01<y<0.1$, which will be available with the detectors at EIC), $\xi$ becomes larger, and as a result, the CFFs in Sec. ~\ref{scattering amp} receive enhanced contributions from valence regions of GPDs.
Another interesting observation from Fig.~\ref{beta_EIC} is that for colliding energy $5\,\text{GeV}\times 41\,\text{GeV}$, the QED contributions become sizable at large $\beta^\prime$. For example, at $\beta^\prime\sim0.9,$ the QED contribution is around $20$ percent of the QCD contribution. For the production of bottom dijets (shown in purple lines), the differential cross sections become zero at large values of $\beta^\prime$. This can be explained through the relation between $Q^2$ and $\beta^\prime$:
$$Q^2=\frac{\beta^\prime(t-M^2)}{\beta^\prime-1}.$$
Since $M^2$ is large for bottom dijets, for large values of $\beta^\prime$, $Q^2$ can easily exceed the considered kinematic range. 
The comparison with HERA data can also be found in ~\cite{Chall:2026oes}, where only the production of quark dijet and contributions from unpolarized GPDs are considered.

\begin{figure}[htbp]
\centering
\includegraphics[width=1.0\textwidth]{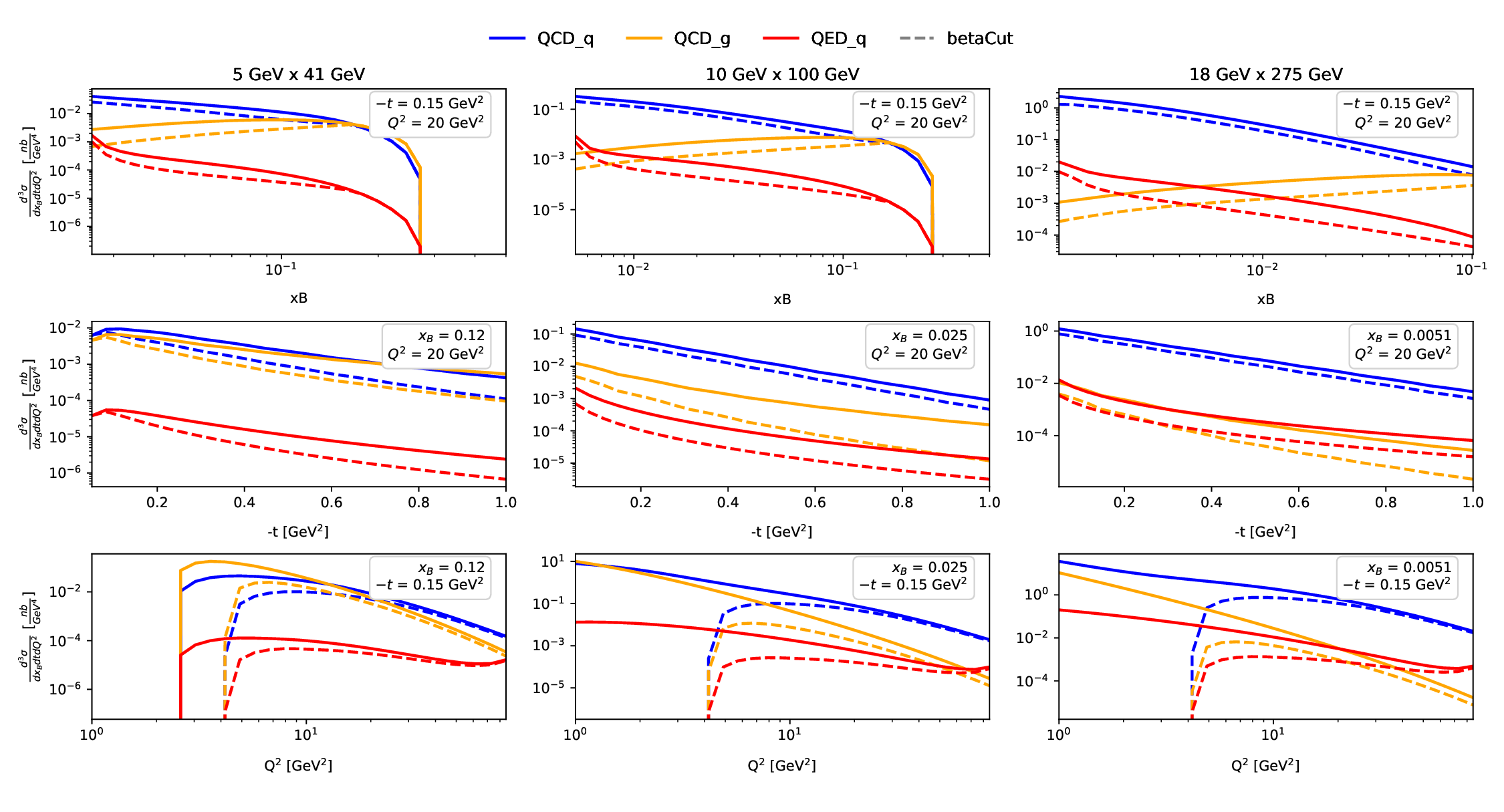}
\caption{The cross section $d^3\sigma/(dx_B\,dt\,dQ^2)$ for different EIC collision energies. For each row, we fix two variables in $(x_B,t,Q^2)$ and show the dependence on the remaining variable. The following kinematic cuts are used in the numerical integration: $1\,\text{GeV}^2<\mathbf{q_\perp^2}<50\,\text{GeV}^2$, $0.05<z<0.95$, $\mu_{\text{F}}^2>1\,\text{GeV}^2.$ The integrations over $\phi$ and $\phi_{p^\prime}$ are performed over the full range.
The blue lines, red lines, and orange lines denote the QCD contributions to $q\bar{q}$ production, QED contributions to $q\bar{q}$ production, and results for $gg$ production, respectively. The solid lines denote the results without a cutoff for $\beta^\prime$, the dashed lines denote the results with a $\beta^\prime$ cutoff: $\beta^\prime\in[0.5,1]$. The correspondences between colors and different contributions remain the same in the latter diagrams.}
\label{xBtQ2EIC}
\end{figure}

Figure~\ref{xBtQ2EIC} displays the differential cross section $d^3\sigma/(dx_B\,dt\,dQ^2)$. For the upper panels, we fix $t$ and $Q^2$ while varying $x_{\text{Bj}}$. A common feature across all colliding energies is that the cross sections for gluon dijet production (shown in orange lines) increase with $x_B$, since $\xi$ increases with $x_B$.  We also investigate the effects of imposing the cut $\beta^\prime\in[0.5,1]$. The results indicate that the contributions from the small-$\beta^\prime$ region become less important with increasing $Q^2$, as illustrated in the bottom panels of Fig.~\ref{xBtQ2EIC}. As $Q^2$ increases, $\beta^\prime\equiv Q^2/(M^2+Q^2-t)$ is forced to be closer to 1.

\begin{figure}[htbp]
\centering
\includegraphics[width=1.0\textwidth]{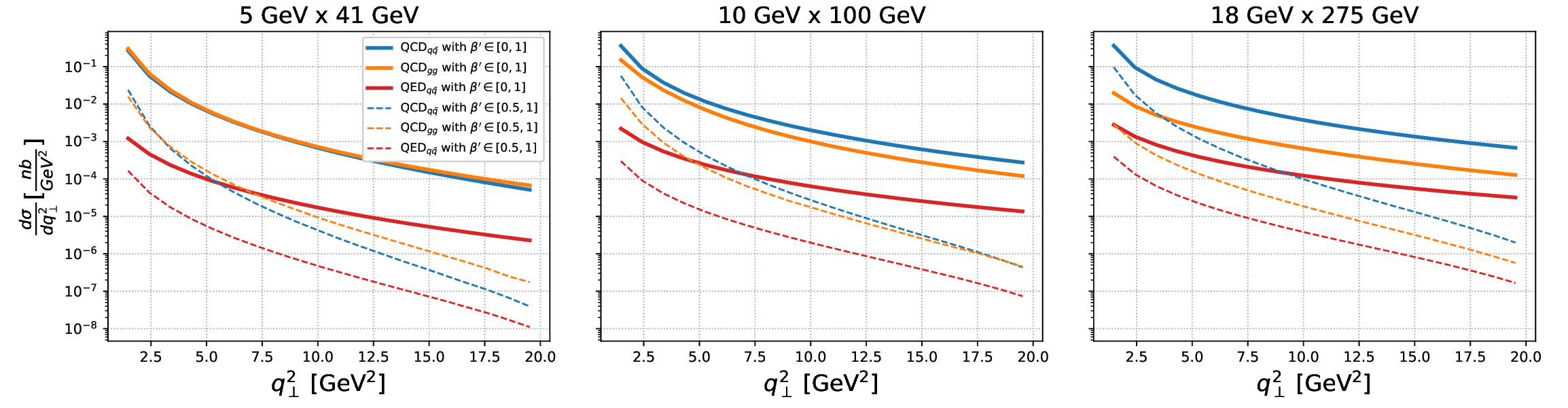}
\caption{The cross section $d\sigma/d\mathbf{q_\perp^2}$ as a function of $\mathbf{q_\perp^2}$ for different EIC collision energies. The following kinematic cuts are used in the numerical integration: $0.01<y<0.95$, $-1.5\,\text{GeV}^2<t<-0.01\,\text{GeV}^2$,  $0.05<z<0.95$, $1\,\text{GeV}^2<Q^2<100\,\text{GeV}^2$. The integrations over $\phi$ and $\phi_{p^\prime}$ are performed over the full range.}
\label{dqperp2}
\end{figure}

The differential cross section $d\sigma/d\mathbf{q_\perp^2}$ is particularly interesting, as shown in Fig.~\ref{dqperp2}. At all colliding energies, the QED contributions become more significant as $\mathbf{q_\perp^2}$ increases, which is a simple reflection that: $$\frac{\big|M_{\text{QED}}^{q\bar{q}}\big|^2}{\big|M_{\text{QCD}}^{q\bar{q}}\big|^2}\propto\frac{\alpha_{\text{em}}^2}{\alpha_s^2(\mu_F^2)}.$$ For higher values of $\mathbf{q_\perp^2}$, $\alpha_s(\mu_F^2)$ becomes smaller. Due to the possible pile-up of particles from different jets, reconstruction of dijets with $\mathbf{q_\perp^2}\sim 1\text{GeV}^2$ is challenging, and whether it can be achieved in the future EIC is worth exploring.

\begin{figure}[htbp]
\centering
\subfloat
{
\includegraphics[width=0.9\textwidth]{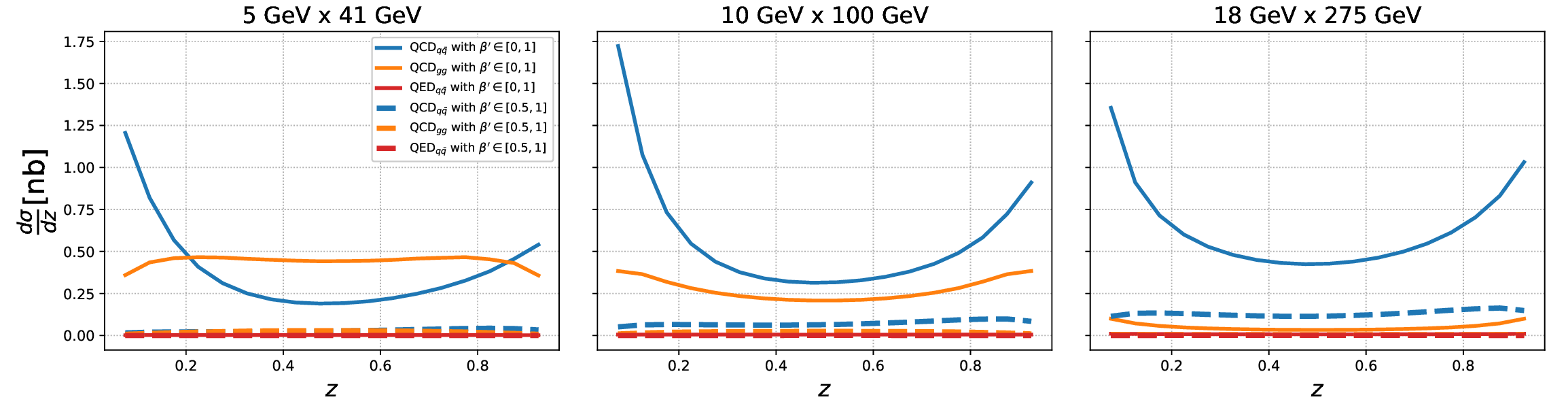}
}
\hfill
\subfloat
{
\includegraphics[width=0.9\textwidth]{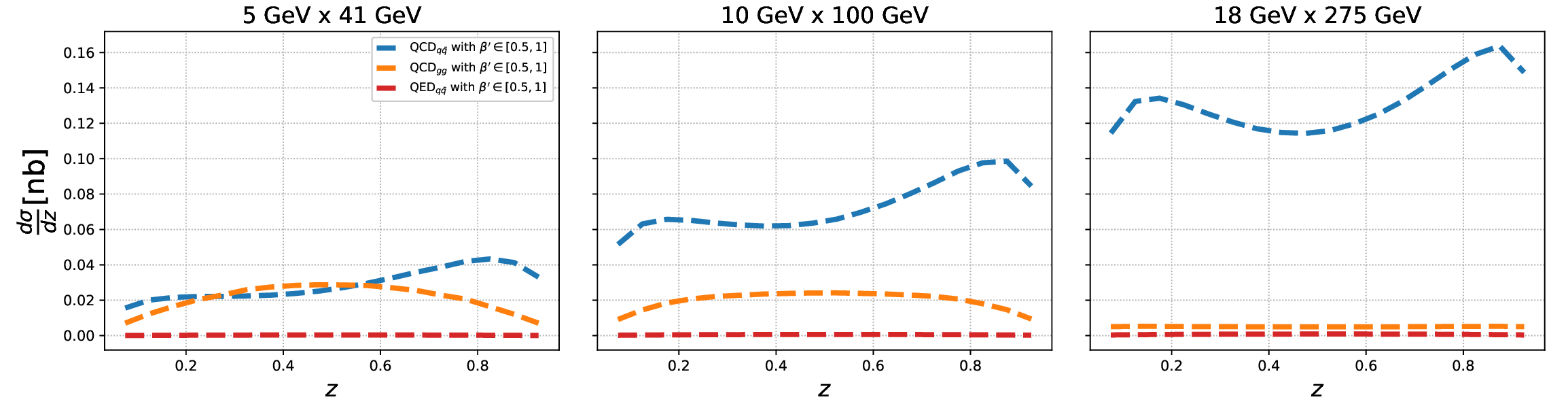}
}
\caption{The cross section $d\sigma/dz$ as a function of $z$ for different EIC collision energies. Upper panels: differential cross sections with and without $\beta^\prime$ cuts, lower panels: differential cross sections with $\beta^\prime$ cuts.  The following kinematic cuts are used in the numerical integration: $0.01<y<0.95$, $-1.5\,\text{GeV}^2<t<-0.01\,\text{GeV}^2$, $1\,\text{GeV}^2<\mathbf{q_\perp^2}<50\,\text{GeV}^2$, $1\,\text{GeV}^2<Q^2<100\,\text{GeV}^2$. The integrations over $\phi$ and $\phi_{p^\prime}$ are performed over the full range.}
\label{zcutfull}
\end{figure}

\begin{figure}[htbp]
\centering
\subfloat
{
\includegraphics[width=0.9\textwidth]{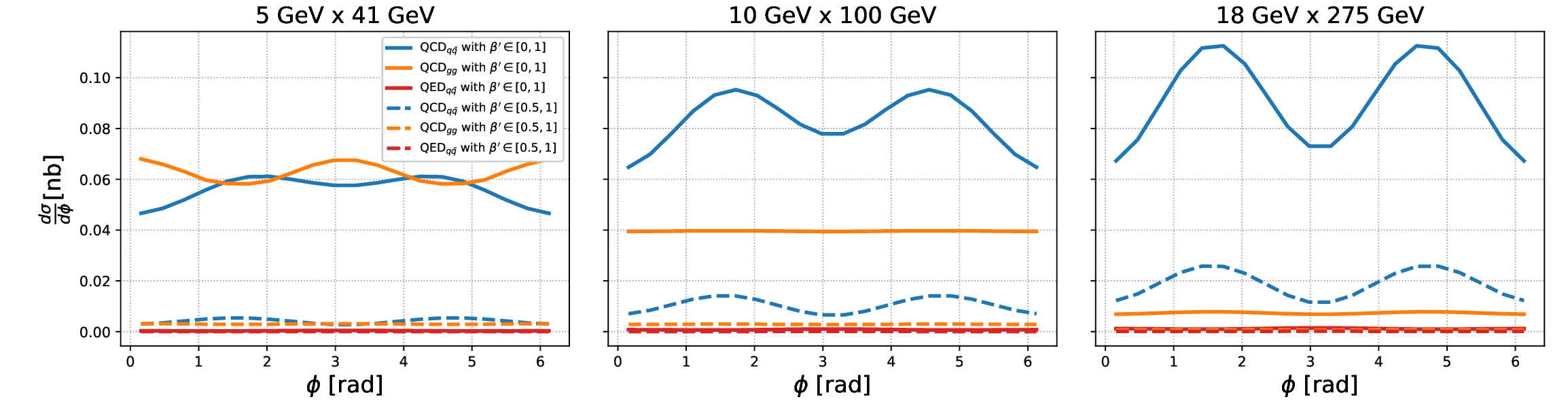}
}
\hfill
\subfloat
{
\includegraphics[width=0.9\textwidth]{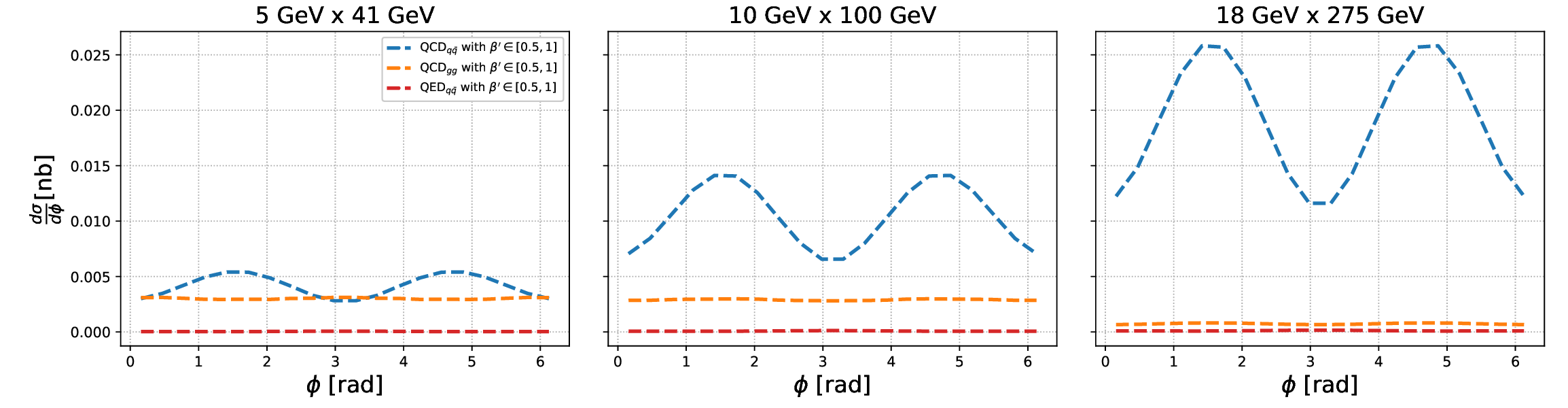}
}
\caption{The cross section $d\sigma/d\phi$ as a function of $\phi$ for different EIC collision energies. Upper panels: differential cross sections with and without $\beta^\prime$ cuts, lower panels: differential cross sections with $\beta^\prime$ cuts. The following kinematic cuts are used in the numerical integration: $0.01<y<0.95$, $-1.5\,\text{GeV}^2<t<-0.01\,\text{GeV}^2$, $1\,\text{GeV}^2<\mathbf{q_\perp^2}<50\,\text{GeV}^2$, $0.05<z<0.95$, $1\,\text{GeV}^2<Q^2<100\,\text{GeV}^2$. The integration $\phi_{p^\prime}$ is performed over the full range.}
\label{phicutfull}
\end{figure}

For exclusive electroproduction of quark dijets, both quark GPDs and gluon GPDs contribute. To illustrate the impacts of valence quark GPDs, we plot the differential cross section $d\sigma/dz$ in Fig.~\ref{zcutfull}.  For quark dijet production, $d\sigma/dz$ will be symmetric with respect to $z=0.5$ if only the sea quark and gluon distributions contribute \footnote{One can prove this from Eqs.~\eqref{qqbaruq} and~\eqref{Igh} by using the specific symmetries of sea quark GPDs and gluon GPDs with respect to $(x\rightarrow-x)$, with contributions from valence quark GPDs neglected. Thus, the obvious derivation of $d\sigma/dz$ from the symmetric shape is a manifestation of the significant contributions from valence quark GPDs.}. In HERA kinematics, the symmetric shapes are observed (see Fig.~5, Fig.~7, Fig.~9, Fig.~11, Fig.~12, and Fig.~14 in ~\cite{Braun:2005rg}). In contrast, due to the lower collision energies of EIC, valence quark GPDs play a more significant role, resulting in an obvious deviation from the symmetric shape in Fig.~\ref{zcutfull}. As shown in ~\cite{Bartels:1996tc} using expansion of the wave function, the quark dijets prefer to be produced in the direction perpendicular to the leptonic plane, when it is the two gluons that are exchanged between the proton and virtual photon. In Fig.~\ref{phicutfull}, we plot the differential cross section $d\sigma/d\phi$ as a function of $\phi$. One interesting observation is that at the lowest colliding energy, the gluon dijet are more likely to be produced when $\phi=0$ or $\phi=\pi$, while at the two higher energies, the production of gluon dijet has no preferred direction. This is due to that the contributions from unpolarized GPDs (Eq.~\eqref{gguu}) and from helicity GPDs (Eq.~\eqref{gghh}) have opposite dependence on $\text{cos}(2\phi)$, which results in nontrivial cancellation.

\vspace{0.5cm}

\begin{figure}[htbp]
\centering
\subfloat
{
\includegraphics[width=1.1\textwidth]{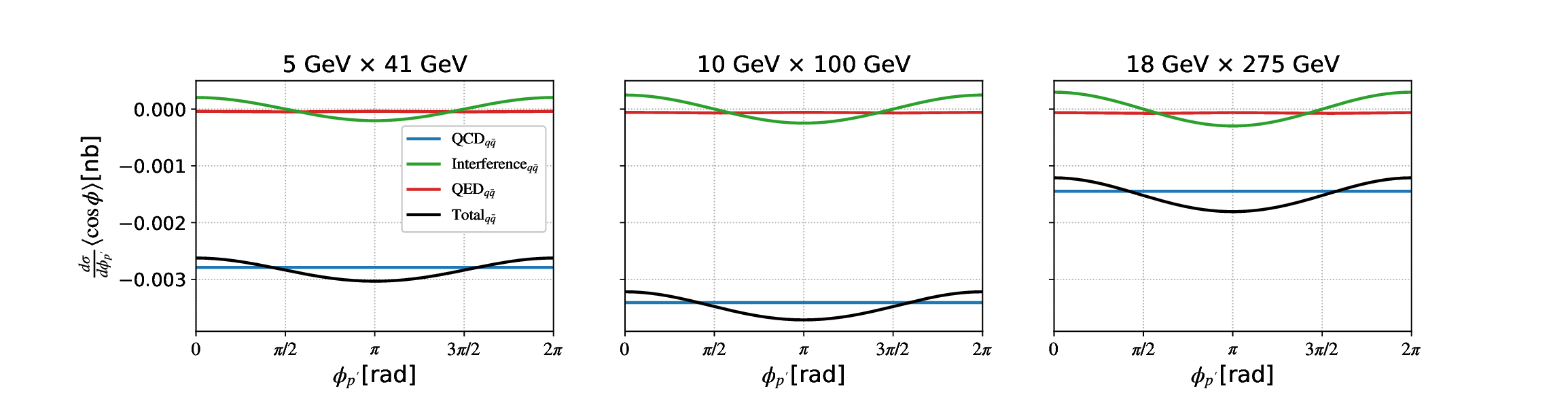}
}
\hfill
\subfloat
{
\includegraphics[width=1.1\textwidth]{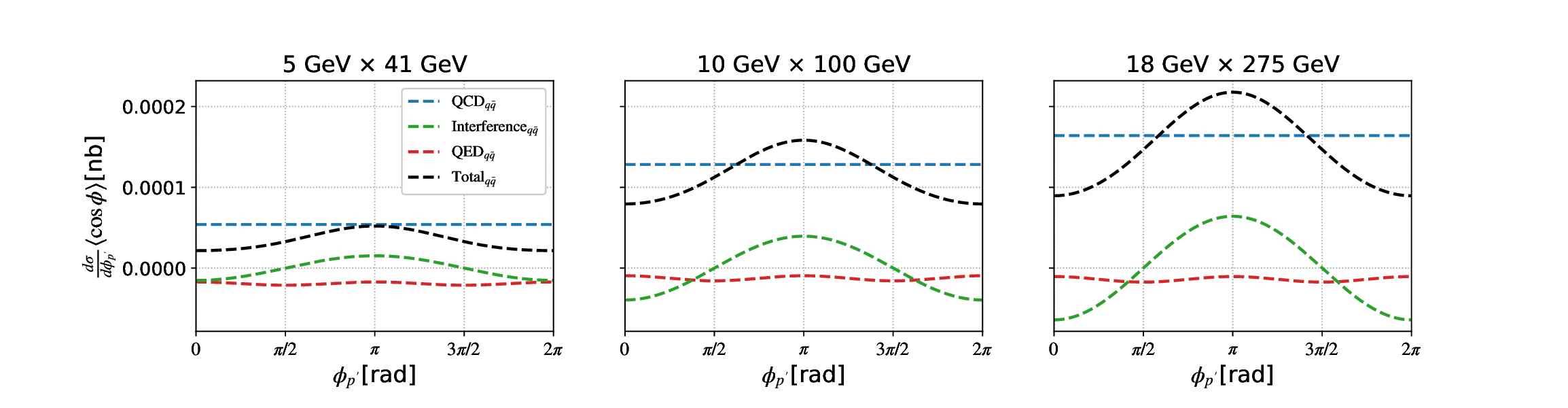}
}
\caption{(Quark dijet production) The cross section $d\sigma/d\phi_{p^\prime}\big<\text{cos}\phi\big>$ as a function of $\phi_{p^\prime}$ for different EIC collision energies. The following kinematic cuts are used in the numerical integration: $0.01<y<0.95$, $0.05<z<0.95$, $-1.5\,\text{GeV}^2<t<-0.01\,\text{GeV}^2$, $1\,\text{GeV}^2<\mathbf{q_\perp^2}<50\,\text{GeV}^2$, $1\,\text{GeV}^2<Q^2<100\,\text{GeV}^2$. The blue, red, and green lines denote the QCD, QED, and interference contributions, respectively.  Upper panels: differential cross sections without $\beta^\prime$ cuts, lower panels: differential cross sections with cut $\beta^\prime\in[0.5,1]$.}
\label{phipinter}
\end{figure}
From Lorentz covariance, the interference contribution depends on $\phi_{p^\prime}$ in the following form:  \begin{equation} \label{interphip}
\text{Re}\big(M_{\text{QCD}}^{q\bar{q}}M_{\text{QED}}^{q\bar{q}*}\big)=A\text{cos}(\phi_{p^\prime})+B\text{sin}(\phi_{p^\prime}),\end{equation}
where $A$ and $B$ are functions of other kinematic variables. 
To visualize the effect of interference contribution, we plot the observable $d\sigma/d\phi_{p^\prime}\big<\text{cos}\phi\big>$ as a function of $\phi_{p^\prime}$ in Fig.~\ref{phipinter}, for quark dijet production. The weighted cross section $d\sigma/d\phi_{p^\prime}\big<\text{cos}\phi\big>$is defined as:
\begin{equation} \label{weighted}
\frac{d\sigma}{d\phi_{p^\prime}}\big<\text{cos}\phi\big>\equiv\int_{0}^{2\pi}\text{cos}\phi\frac{d^2\sigma}{d\phi d\phi_{p^\prime}},
\end{equation}
where $d^2\sigma/(d\phi d\phi_{p^\prime})$ on the r.h.s. can be obtained from Eq.~\eqref{master} by integrating over variables $(y,z,Q^2,\mathbf{q_\perp^2},\mathbf{\Delta_\perp^2},\phi_{l^\prime}).$
Eq.~\eqref{weighted} is sensitive to the interference between amplitudes involving a longitudinally polarized photon and a transversely polarized photon (see Eq.~\eqref{csQCDuu}, Eq.~\ref{csQCDhh}, and Eq.~\eqref{gguu}). 
From Fig.~\ref{phipinter}, one can observe that the total cross sections receive sizable contributions from the QED channel, especially for the lower panels.

For some observables presented in this section, the QED contributions become more significant in the larger $\beta^\prime$ region ($\beta^\prime\gtrsim 0.5$, see Fig.~\ref{beta_EIC}, Fig.~\ref{dqperp2}, and Fig.~\ref{phipinter}). This fact can be explained by the different behaviors of QCD and QED contributions at small $\mathbf{q_\perp^2}$ and large $Q^2$ region (discussed in Sec.~\ref{scattering amp}).
To obtain a large $\beta^\prime$, one requires either a small (still perturbative) $\mathbf{q_\perp^2}$, a large $Q^2$, or both.

\section{Conclusions}
\label{con}
In this work, we consider the exclusive electroproduction of quark and gluon dijets. Such a process is sensitive to generalized parton distributions. We derive the leading-order expressions for the production of quark and gluon dijets, at amplitude and cross section level. While NLO corrections have proved to be important in many exclusive processes, the systematic study of the leading-order contributions from different channels presented here constitutes an important first step toward a detailed NLO analysis of quark- and gluon-dijet production. We find that the QED contributions are sizable in certain observables when $\beta^\prime$ is large.
Compared with HERA kinematics, the cross sections receive enhanced contributions from valence quark GPDs in EIC kinematics.  As a consequence, the production of gluon dijets becomes more pronounced in EIC kinematics.
In our phenomenology analyses, we present the results for quark and gluon dijet production separately, as these two processes probe different types of GPDs. At future facilities such as EIC and EicC, the ability to discriminate between quark- and gluon-initiated jets will play a crucial role and is expected to benefit from ongoing developments in jet-tagging techniques~\cite{Qu:2019gqs}. While exclusive electroproduction of heavy-flavor quark dijets probes gluon GPDs exclusively, the corresponding light-flavor dijets production receives contributions from both quark and gluon GPDs. A further separation of heavy- and light-flavor jets relies on the precise reconstruction of secondary decay vertices; recent progress in this area can be found in ~\cite{Guest:2016iqz,CMS:2017wtu,Qu:2019gqs}. 

In addition to the case when beam and target are both unpolarized as considered in this work, one can also consider the polarized target, polarized beam, or both. Although further investigation is required to verify whether the cross-section is sufficiently large, extending our study to proton-to-neutron transition GPDs~\cite{Mankiewicz:1997aa,Duplancic:2022ffo} could provide a unique tool to isolate quark contributions, since in this case gluon distributions, gluon jets, and the Bethe-Heitler background do not contribute. A realistic description of experimentally observed jets requires connecting the produced quark--antiquark and gluon--gluon pairs at the partonic level to the corresponding hadronic final states. This necessitates the inclusion of hadronization effects. We leave these studies for future publications.

\begin{acknowledgments}
We thank Markus Diehl for the comments on the first-order derivatives of gluon GPDs. We also thank Valerio Bertone, Cédric Mezrag, Maxim Nefedov, Bernard Pire, Wolfgang Schäfer, and Antoni Szczurek for valuable discussions and comments. This research was funded in whole or in part by the National Science Centre, Poland (grant OPUS 27 No.~2024/53/B/ST2/00968). For the purpose of open access, the authors have applied a CC-BY public copyright licence to any Author Accepted Manuscript (AAM) version arising from this submission.
\end{acknowledgments}
\clearpage
\appendix
\section{Analysis of leading-order evolution of gluon GPDs}
In this appendix, we show that the first derivatives of gluon GPDs are continuous under evolution, which is crucial for the existence of factorization considered in this work.
We start from the standard evolution of gluon GPDs:
\begin{align} \label{evo_general}
\frac{d}{d\text{ln}\mu}F_{gi}(x,\xi,t)=\int_{-1}^1dy\Big[K^{gg}_i(x_1,x_2\big|y_1,y_2)F_{gi}(y,\xi,t)+K^{gq}_i(x_1,x_2\big|y_1,y_2)F_{qi}^{\text{s}}(y,\xi,t)\Big],
\end{align}
where $x_1=\frac{\xi+x}{2},x_2=\frac{\xi-x}{2},y_1=\frac{\xi+y}{2},y_2=\frac{\xi-y}{2}$, the superscript ``s" denotes the singlet sector. The subscript $``i"$ denotes the specific type of GPDs: $$i\in\{\text{unpolarized}(u),\text{helicity}(h),\text{transversity}(T)\}.$$ $K^{gg}(K^{gq})$ is the gluon-in-gluon (gluon-in-quark) evolution kernel, the specific expressions can be found in ~\cite{Diehl:2003ny,Belitsky:2005qn} (we have $K_{T}^{gq}=0$). We suppress the dependence on $t$ in the discussions below. The following shorthand notations for an arbitrary function $f(x,y)$ will be used:
\begin{equation}
\partial_xF(x,y)\equiv\frac{\partial f(x,y)}{\partial x},\quad\partial_yF(x,y)\equiv\frac{\partial f(x,y)}{\partial y}.
\end{equation}
We are interested in the following difference:
\begin{equation} 
\left(\frac{d(\partial_xF_{gi}(\xi,\xi))}{d\text{ln}\mu}\right)^{\pm}\equiv\lim_{x\to\xi^+}\Big(\frac{d(\partial_xF_{gi}(x,\xi))}{d\text{ln}\mu}\Big)-\lim_{x\to\xi^-}\Big(\frac{d(\partial_xF_{gi}(x,\xi))}{d\text{ln}\mu}\Big),
\end{equation}
where $\lim_{x\to\xi^{+}/\xi^-}\Big(\frac{d(\partial_xF_{gi}(x,\xi))}{d\text{ln}\mu}\Big)$ is understood as first taking the derivative of (the r.h.s. of) Eq.~\eqref{evo_general} with respect to $x$ in the region $x>\xi/x<\xi$, then taking the limit $x\rightarrow\xi$ of the results from upper side/lower side. We assume at scale $\mu^2$, $\partial_xF_{gi}(x,\xi)$ is $\mathcal{C}^1$ in the region $x\in(-1,1)$, and $\partial^2 F_{gi}(x,\xi)/\partial^2x$ is finite (can be discontinuous) in the neighbourhood of $x=\xi$ for all $i$. 

\subsection{Gluon-in-gluon channel}
The gluon-in-gluon channel takes the following form ~\cite{Belitsky:2005qn}:
\begin{align} \label{gg_initial}
\frac{d}{d\text{ln}\mu}F_{gi}(x,\xi)=&\int_{-1}^1dy\Big\{\frac{x_1}{y_1}\Big[\frac{x_1}{x_1-y_1}\vartheta_{11}^0(x_1,x_1-y_1)\Big]_++\frac{x_2}{y_2}\Big[\frac{x_2}{x_2-y_2}\vartheta_{11}^0(x_2,x_2-y_2)\Big]_+\nonumber \\
&+\Delta K^{gg}_i(x_1,x_2\big|y_1,y_2)+\Big(\frac{1}{2}\frac{\beta_0}{C_A}+2\Big)\delta(x-y)\Big\}F_{gi}(y,\xi),
\end{align}
the plus distribution is defined as:
\begin{equation}
\Big[\frac{x_1}{x_1-y_1}\vartheta_{11}^0(x_1,x_1-y_1)\Big]_+\equiv\frac{x_1}{x_1-y_1}\vartheta_{11}^0(x_1,x_1-y_1)-\delta(x_1-y_1)\int dx_1^\prime\frac{x_1^\prime}{x_1^\prime-y_1}\vartheta_{11}^0(x_1^\prime,x_1^\prime-y_1).
\end{equation}
The $\Delta K_i^{gg}(x_1,x_2\big|y_1,y_2)$ in the fourth line depends on the specific type of GPDs~\cite{Belitsky:2005qn}:
\begin{equation} \label{DeltaKgg}
\Delta K_{i}^{gg}(x,y,\xi)=
\begin{cases}
2\,\dfrac{x_1x_2+y_1y_2}{y_1y_2}\,
\vartheta_{111}^{0}(x_1,-x_2,x_1-y_1)
+2\,\dfrac{x_1x_2(x_1y_1+x_2y_2)}{y_1y_2\,(x_1+x_2)^2}\,
\vartheta_{11}^{0}(x_1,-x_2),
& i=u, \\[1.2ex]
2\,\dfrac{x_1y_2+y_1x_2}{y_1y_2}\,
\vartheta_{111}^{0}(x_1,-x_2,x_1-y_1)
+2\,\dfrac{x_1x_2}{y_1y_2}\,
\vartheta_{11}^{0}(x_1,-x_2),
& i=h, \\[1.2ex]
0, & i=t.
\end{cases}
\end{equation}
the definition of $\vartheta_{11}^0$ and $\vartheta_{111}^0$ can be found in Appendix. G7 of ~\cite{Belitsky:2005qn}.

Our analyses below are based on an equivalent form of Eq.~\eqref{gg_initial}:
\begin{align} \label{gg_general}
\frac{d}{d\text{ln}\mu}F_{gi}(x,\xi)=&\int_{-1}^1dy\Big\{\Big[\frac{x_1}{y_1}\frac{x_1}{x_1-y_1}\vartheta_{11}^0(x_1,x_1-y_1)+\frac{x_2}{y_2}\frac{x_2}{x_2-y_2}\vartheta_{11}^0(x_2,x_2-y_2)\Big]\big(F_{gi}(y,\xi)-F_{gi}(x,\xi)\big)\nonumber \\
&+\Big[\frac{x_1}{y_1}\frac{x_1}{x_1-y_1}\vartheta_{11}^0(x_1,x_1-y_1)+\frac{x_2}{y_2}\frac{x_2}{x_2-y_2}\vartheta_{11}^0(x_2,x_2-y_2)-\frac{y_1}{y_1-x_1}\vartheta_{11}^0(y_1,y_1-x_1)\nonumber \\
&-\frac{y_2}{y_2-x_2}\vartheta_{11}^0(y_2,y_2-x_2)\Big]F_{gi}(x,\xi)\nonumber \\
&+\Delta K^{gg}_i(x_1,x_2\big|y_1,y_2)+\Big(\frac{1}{2}\frac{\beta_0}{C_A}+2\Big)\delta(x-y)\Big\}F_{gi}(y,\xi),
\end{align}

\subsubsection{First line of Eq.~\texorpdfstring{\eqref{gg_general}}{(\ref*{gg_general})}}
For the first line of Eq.~\eqref{gg_general}, we have ($G(x,y)=(F_{gi}(y,\xi)-F_{gi}(x,\xi))/(x-y)$):
\begin{align} \label{first_line}
\left(\frac{d(\partial_xF_{gi}(\xi,\xi))}{d\text{ln}\mu}\right)^{\pm}\Big|_{gg,\text{1st line}}
=&\lim_{x\to\xi^+}\Big(4(x-\xi)\int_{x}^1\frac{G(x,y)}{(y-\xi)^2}dy+2\cdot\partial_xF_{gi}(x,\xi)+2(x-\xi)^2\int_x^1\frac{\partial_xG(x,y)}{(y-\xi)^2}dy\Big)\nonumber \\
&-\lim_{x\to\xi^-}\Big(-4(x-\xi)\int_{-1}^x\frac{G(x,y)}{(y-\xi)^2}dy+2\cdot\partial_xF_{gi}(x,\xi)-2(x-\xi)^2\int_{-1}^x\frac{\partial_xG(x,y)}{(y-\xi)^2}dy\Big),
\end{align}
the subscript $``gg"$ denotes the gluon-in-gluon channel. Note that due to the existence of $\partial^2F(x,\xi)/\partial^2x$ in the neighbourhood of $x=\xi$, both $\partial_xG(x,y)$ and $\partial_y G(x,y)$ are bounded from above in the region $y\in[-1,1]$ ($C$ and $C^\prime$ are some finite constants):
$$\partial_xG(x,y)\leq C,$$
$$\partial_yG(x,y)\leq C^\prime.$$
Which allows us to write the first and second lines of Eq.~\eqref{first_line} as:
\begin{align}
&\lim_{x\rightarrow\xi^+}\Big(4(x-\xi)\int_{x}^1\frac{G(x,y)}{(y-\xi)^2}dy+2\cdot\partial_xF_{gi}(x,\xi)+2(x-\xi)^2\int_x^1\frac{\partial_xG(x,y)}{(y-\xi)^2}dy\Big)\nonumber \\
=&\lim_{x\rightarrow\xi^+}\left(4(x-\xi)\left(
\int_x^1 \frac{G(x,y)-G(x,\xi)}{(y-\xi)^2}\,dy
+\int_x^1 \frac{G(x,\xi)}{(y-\xi)^2}\,dy
\right)+2\cdot\partial_xF_{gi}(x,\xi)\right)\nonumber \\
=&\lim_{x\to \xi^+}\left( 4(x-\xi)\left(
\int_x^1 \partial_y G(x,y)\,\frac{1}{(y-\xi)}\,dy
+ G(x,\xi)\left(\frac{1}{x-\xi}-\frac{1}{1-\xi}\right)
\right)+2\cdot\partial_xF_{gi}(x,\xi)\right)\nonumber \\
=&-2\partial_xF_{gi}(\xi^+,\xi),\nonumber \\
&-\lim_{x\rightarrow\xi^-}\Big(-4(x-\xi)\int_{-1}^x\frac{G(x,y)}{(y-\xi)^2}dy+2\cdot\partial_xF_{gi}(x,\xi)-2(x-\xi)^2\int_{-1}^x\frac{\partial_xG(x,y)}{(y-\xi)^2}dy\Big)\nonumber \\
=&2\partial_xF_{gi}(\xi^-,\xi),
\end{align}
where from the second line to the third line, we have used integration by parts. In the limit $x\rightarrow\xi^{\pm}$ the boundary terms vanish when multiplied with the $(x-\xi)$ factor. Combine everything together, we have:
\begin{align}
\left(\frac{d(\partial_xF_{gi}(\xi,\xi))}{d\text{ln}\mu}\right)^{\pm}\Big|_{gg,\text{1st line}}=-2\partial_xF_{gi}(\xi^+,\xi)+2\partial_xF_{gi}(\xi^-,\xi)=0.
\end{align}
\subsubsection{Second and third lines of Eq.~\texorpdfstring{\eqref{gg_general}}{(\ref*{gg_general})}}
The integration in the second and third lines can be performed, which gives:
\begin{equation} \label{2nd and 3rd lines}
\frac{d}{d\ln\mu} F_{gi}^{gg}(x, \xi)\big|_{gg,\text{2nd and 3rd lines}}= F_{gi}(x, \xi) \cdot 
\begin{cases}
2\ln \left(1-\xi^2\right) + \frac{4\left(x-\xi^2\right)}{\xi^2-1} - 4\ln(1-x), & x > \xi \\
-\frac{4\xi}{1+\xi} + 4\ln(\xi+1) - 2\ln \left(1-x^2\right), & 0 < x < \xi
\end{cases}
\end{equation}
where the principal-value prescription is used when $x>\xi$. From Eq.~\eqref{2nd and 3rd lines}, We find:
\begin{align}
&\lim_{x\to\xi^+}\Big(\frac{d(\partial_xF_{gi}(x,\xi))}{d\text{ln}\mu}\Big)\big|_{gg,\text{2nd and 3rd lines}}\nonumber \\
=&\partial_xF_{gi}(\xi^+,\xi)\cdot\big(\frac{-4\xi}{1+\xi}+2\text{ln}\frac{1+\xi}{1-\xi}\big)+F_{gi}(\xi,\xi)\cdot\frac{4\xi}{1-\xi^2}\nonumber \\
=&\lim_{x\to\xi^-}\Big(\frac{d(\partial_xF_{gi}(x,\xi))}{d\text{ln}\mu}\Big)\big|_{gg,\text{2nd and 3rd lines}},
\end{align}
thus
\begin{equation} \left(\frac{d(\partial_xF_{gi}(\xi,\xi))}{d\text{ln}\mu}\right)^{\pm}\Big|_{gg,\text{2nd and 3rd lines}}=0.
\end{equation}

\subsubsection{Fourth line of Eq.~\texorpdfstring{\eqref{gg_general}}{(\ref*{gg_general})}} \label{gg_gq}
We can rewrite the fourth line of Eq.~\eqref{gg_general} as:
\begin{equation} \label{DeltaKgg1}
\frac{d}{d\ln\mu} F_{gi}^{gg}(x, \xi)\big|_{gg,\text{4th line}}=
\begin{cases}
  \int_{x}^1dy\Delta K^{gg}_i(x,y,\xi)\big|_{y>x,x>\xi}F_{gi}(y,\xi)+\int_{-1}^xdy\Delta K^{gg}_i(x,y,\xi)\big|_{x>y,x>\xi}F_{gi}(y,\xi)\\
  +\Big(\frac{1}{2}\frac{\beta_0}{C_A}+2\Big)F_{gi}(x,\xi),& x>\xi \\
  \int_{x}^1dy\Delta K^{gg}_i(x,y,\xi)\big|_{y>x,x<\xi}F_{gi}(y,\xi)+\int_{-1}^xdy\Delta K^{gg}_i(x,y,\xi)\big|_{x>y,x<\xi}F_{gi}(y,\xi)\\ 
  +\Big(\frac{1}{2}\frac{\beta_0}{C_A}+2\Big)F_{gi}(x,\xi),& x<\xi
\end{cases}
\end{equation}
the specific form of $\Delta K^{gg}_i(x,y,\xi)\big|_{x\gtrless y,x\gtrless\xi}$ can be calculated from Eq.~\eqref{DeltaKgg} by taking $x\gtrless y$ and $x\gtrless\xi$. Taking derivative of Eq.~\eqref{DeltaKgg1} with respect to $x$, we find (note that in Eq.~\eqref{DeltaKgg1}, both the integrand and integration limit depend on $x$):
\begin{equation}
\left(\frac{d(\partial_xF_{gi}(\xi,\xi))}{d\text{ln}\mu}\right)^{\pm}\Big|_{gg,\text{4th line}}=0,
\end{equation}
for all types of gluon GPDs ($i=u,h,T$).

\subsection{Gluon-in-quark channel}
The leading-order gluon-in-quark evolution kernels in Eq.~\eqref{evo_general} take the following form ~\cite{Belitsky:2005qn}:
\begin{align}
K^{gq}_u(x_1,x_2\big|y_1,y_2)=&C_F\big[(y_1-y_2)\vartheta_{111}^0(x_1,-x_2,x_1-y_1)+x_1x_2\vartheta_{111}^1(x_1,-x_2,x_1-y_1)\big],\nonumber \\
K^{gq}_h(x_1,x_2\big|y_1,y_2)=&C_F\big[(x_1-x_2)\vartheta_{111}^0(x_1,-x_2,x_1-y_1)+x_1x_2\vartheta_{111}^1(x_1,-x_2,x_1-y_1)\big],
\end{align}
the definition of $\vartheta_{111}^1(x_1,-x_2,x_1-y_1)$ can be found in Appendix. G7 of ~\cite{Belitsky:2005qn}. Following the treatment in Sec. ~\ref{gg_gq}, we find:
\begin{align} \label{gq_diff}
\left(\frac{d(\partial_xF_{gu}(\xi,\xi))}{d\text{ln}\mu}\right)^{\pm}\Big|_{gq}=&\int_{-1}^1dy\frac{1}{\xi}F_{qu}^{\text{s}}(y,\xi),\nonumber \\
\left(\frac{d(\partial_xF_{gh}(\xi,\xi))}{d\text{ln}\mu}\right)^{\pm}\Big|_{gq}=&0.
\end{align}
the subscript ``gq" denotes the gluon-in-quark channel. The r.h.s. of the first equation of Eq.~\eqref{gq_diff} vanishes due to symmetry: $\frac{1}{\xi}$ is $\mathbf{even}$ in $(y\rightarrow-y)$ while $F_{qu}^{\text{s}}(y,\xi)$ is $\mathbf{odd}$ in $(y\rightarrow-y)$. 

\subsection{Conclusion}
Combining the above results, we find that the evolution of gluon GPDs won't generate discontinuities for its first derivative at $x=\xi$:
\begin{equation}
\left(\frac{d(\partial_xF_{gi}(\xi,\xi))}{d\text{ln}\mu}\right)^{\pm}=0,
\end{equation}
for all types of GPDs ($i=u,h,T$). The above analyses can be similarly performed for the point $x=-\xi$, with the assumption that $\partial^2 F_{gi}(x,\xi)/\partial^2x$ is finite (can be discontinuous) in the neighbourhood of $x=-\xi$ for all $i$, one finds:
\begin{equation}
\left(\frac{d(\partial_xF_{gi}(-\xi,\xi))}{d\text{ln}\mu}\right)^{\pm}=0.
\end{equation}
\newpage
\section{Analytical results of \texorpdfstring{$\big|M_{\text{QED}}\big|^2$}{Pure QED}
and
\texorpdfstring{$2\mathrm{Re}(M_{\text{QCD}}M_{\text{QED}}^*)$}{interference}}

\subsection{\texorpdfstring{$2\text{Re}\big(M^{q\bar{q}}_{\text{QCD}}M^{q\bar{q}*}_{\text{QED}}\big)$}{interference}}\label{inter_form}
We have $$2\text{Re}\big(M^{q\bar{q}}_{\text{QCD}}M^{q\bar{q}*}_{\text{QED}}\big)=2\text{Re}\big(M^{q\bar{q},u}_{\text{QCD}}M^{q\bar{q}*}_{\text{QED}}\big)+2\text{Re}\big(M^{q\bar{q},h}_{\text{QCD}}M^{q\bar{q}*}_{\text{QED}}\big),$$ note that due to the existence of $J_\perp^\alpha$ in $M_{QED}^{q\bar{q}}$, $\text{Re}\big(M^{q\bar{q},h}_{\text{QCD}}M^{q\bar{q}*}_{\text{QED}}\big)$ is not zero for the unpolarized target.
Their results read :
\begin{align} \label{priQCDu}
2\text{Re}\big(M^{q\bar{q},u}_{\text{QCD}}M^{q\bar{q}*}_{\text{QED}}\big)=&-\frac{128 \pi ^4 \alpha_{\text{em}}^3\alpha_s e_q^2 }{Q^2 t \left(\mu^2+\mathbf{q_\perp^2}\right)^2}\Big\{\text{Re}\Big[(2C_FI_{qu1}^{q\bar{q}}+I_{gu1})\big((\Sigma_fe_f\tilde{J}_f^{x*})\cdot\Sigma_{i=0}^2a_{gqu1,x}^i\text{cos}(i\phi)\nonumber \\
&+(\Sigma_fe_f\tilde{J}_f^{y*})\cdot\Sigma_{i=1}^2a_{gqu1,y}^i\text{sin}(i\phi)\big)\Big]-\text{Re}\Big[(4C_FI_{qu2}^{q\bar{q}}+I_{gu2})\big((\Sigma_fe_f\tilde{J}_f^{x*})\cdot\Sigma_{i=0}^3a_{gqu2,x}^i\text{cos}(i\phi)\nonumber \\
&+(\Sigma_fe_f\tilde{J}_f^{y*})\cdot\Sigma_{i=1}^3a_{gqu2,y}^i\text{sin}(i\phi)\big)\Big]-\text{Re}\Big[(4C_FI_{qu3}^{q\bar{q}}-I_{gu2})\big((\Sigma_fe_f\tilde{J}_f^{x*})\cdot\Sigma_{i=0}^3a_{gqu3,x}^i\text{cos}(i\phi)\nonumber \\
&+(\Sigma_fe_f\tilde{J}_f^{y*})\cdot\Sigma_{i=1}^3a_{gqu3,y}^i\text{sin}(i\phi)\big)\Big]\Big\},
\end{align}

the harmonic coefficients in Eq.~\eqref{priQCDu} read:
\begin{align*} \label{priQCDucoeff}
a_{gqu1,x}^0=&-\frac{16 Q \sqrt{1-y} \left(e_q \mu ^2 (-2 y z+y+4 z-2)+8 Q^2 y z^2 \bar{z}^2\right)}{y^2},\nonumber \\
a_{gqu1,x}^1=&-\frac{16 Q^2 \big|q_\perp\big| z \bar{z} \left(8 e_q (y-1) \left(m_q^2+\mathbf{q_\perp^2}\right)-(y-2) y (2 z-1) \left(m_q^2-Q^2 z \bar{z}+\mathbf{q_\perp^2}\right)\right)}{y^2 \left(m_q^2+\mathbf{q_\perp^2}\right)},\nonumber \\
a_{gqu1,x}^2=&-\frac{16 e_q Q \mathbf{q_\perp^2} \sqrt{1-y} (y-2) (2 z-1) }{y^2},\nonumber \\
a_{gqu1,y}^1=&a_{gqu1,x}^1+\frac{32 Q^4 \big|q_\perp\big| (y-2) z^2 (2 z-1) \bar{z}^2 }{y \left(m_q^2+\mathbf{q_\perp^2}\right)},\nonumber \\
a_{gqu1,y}^2=&a_{gqu1,x}^2,\nonumber \\
a_{gqu2,x}^0=&\frac{4 Q \mathbf{q_\perp^2} \bar{z}  \left(2 e_q (y-2) (y-1) \left(m_q^2+\mathbf{q_\perp^2}\right)+((y-2) y+2) y \bar{z} \left(m_q^2-Q^2 z \bar{z}+\mathbf{q_\perp^2}\right)\right)}{\sqrt{1-y} y^2 \left(m_q^2+\mathbf{q_\perp^2}\right)},\nonumber \\
a_{gqu2,x}^1=&-\frac{4 \big|q_\perp\big| \left(e_q \mu^2\left(-y^2+(y-2)^2 z+2 y-2\right)+e_q \bar{z} \mathbf{q_\perp^2} ((y-2) y+2)-4 Q^2 (y-2) y z^2 \bar{z}^2\right)}{y^2 z},\nonumber \\
a_{gqu2,x}^2=&\frac{8 Q \mathbf{q_\perp^2} \sqrt{1-y} \left(e_q (y-2) (z-1) \left(m_q^2+\mathbf{q_\perp^2}\right)+y z \bar{z} \left(-m_q^2+Q^2 z \bar{z}-\mathbf{q_\perp^2}\right)\right)}{y^2 \left(m_q^2+\mathbf{q_\perp^2}\right)},\nonumber \\
a_{gqu2,x}^3=&-\frac{8 e_q \big|q_\perp\big|\mathbf{q_\perp^2} (y-1) }{y^2},\nonumber \\
a_{gqu2,y}^1=&a_{gqu2,x}^1-\frac{16 \big|q_\perp\big|  \left(e_q \mu ^2 (y-1)+Q^2 (y-2) y z \bar{z}^2\right)}{y^2},\nonumber \\
a_{gqu2,y}^2=&a_{gqu2,x}^2-\frac{16 Q^3 \mathbf{q_\perp^2} \sqrt{1-y} z^2 \bar{z}^2}{y \left(m_q^2+\mathbf{q_\perp^2}\right)},\nonumber \\
a_{gqu2,y}^3=&a_{gqu2,x}^3,
\end{align*}
\begin{align}
a_{gqu3,x}^0=&\frac{4 Q \mathbf{q_\perp^2} z \left(2 e_q (y-2) (y-1) \left(m_q^2+\mathbf{q_\perp^2}\right)-y ((y-2) y+2) z \left(m_q^2-Q^2 z \bar{z}+\mathbf{q_\perp^2}\right)\right)}{\sqrt{1-y} y^2 \left(m_q^2+\mathbf{q_\perp^2}\right)},\nonumber \\
a_{gqu3,x}^1=&-\frac{4 \big|q_\perp\big| \left(e_q \left(\mu^2 \left((y-2)^2 z+2 (y-1)\right)-\mathbf{q_\perp^2} ((y-2) y+2) z\right)-4 Q^2 (y-2) y z^2 \bar{z}^2\right)}{y^2 \bar{z}},\nonumber \\
a_{gqu3,x}^2=&-a_{gqu2,x}^2-\frac{8 e_q Q \mathbf{q_\perp^2} \sqrt{1-y} (y-2) }{y^2},\nonumber \\
a_{gqu3,x}^3=&-a_{gqu2,x}^3,\nonumber \\
a_{gqu3,y}^1=&a_{gqu3,x}^1-\frac{16 \big|q_\perp\big| \left(e_q \mu ^2 (1-y)\bar{z}+Q^2 (y-2) y z^2 \bar{z}^2\right)}{y^2 \bar{z}},\nonumber \\
a_{gqu3,y}^2=&a_{gqu3,x}^2+\frac{16 Q^3 \mathbf{q_\perp^2} \sqrt{1-y} z^2 \bar{z}^2 }{y \left(m_q^2+\mathbf{q_\perp^2}\right)},\nonumber \\
a_{gqu3,y}^3=&-a_{gqu2,x}^3.
\end{align}

\begin{align} \label{priQCDh}
2\text{Re}\big(M^{q\bar{q},h}_{\text{QCD}}M^{q\bar{q}*}_{\text{QED}}\big)=&\frac{128 \pi ^4 \alpha_{\text{em}}^3\alpha_s e_q^2 }{Q^2 t \left(\mu^2+\mathbf{q_\perp^2}\right)^2}\Big\{\text{Im}\Big[\big((\Sigma_fe_f\tilde{J}_f^{y})\cdot\Sigma_{i=0}^2a_{qh1,y}^i\text{cos}(i\phi)+(\Sigma_fe_f\tilde{J}_f^{x})\cdot\Sigma_{i=1}^2a_{qh1,x}^i\text{sin}(i\phi)\big)C_FI_{qh1}^{q\bar{q}*}\Big]\nonumber \\
&+\text{Im}\Big[\big((\Sigma_fe_f\tilde{J}_f^{y})\cdot\Sigma_{i=0}^3a_{qh2,y}^i\text{cos}(i\phi)+(\Sigma_fe_f\tilde{J}_f^{x})\cdot\Sigma_{i=1}^3a_{qh2,x}^i\text{sin}(i\phi)\big)C_FI_{qh2}^{q\bar{q}*}\Big]\nonumber \\
&+\text{Im}\Big[\big((\Sigma_fe_f\tilde{J}_f^{y})\cdot\Sigma_{i=0}^3a_{qh3,y}^i\text{cos}(i\phi)+(\Sigma_fe_f\tilde{J}_f^{x})\cdot\Sigma_{i=1}^3a_{qh3,x}^i\text{sin}(i\phi)\big)C_FI_{qh3}^{q\bar{q}*}\Big]\nonumber \\
&+\text{Im}\Big[\big((\Sigma_fe_f\tilde{J}_f^{y})\cdot\Sigma_{i=0}^3a_{gh,y}^i\text{cos}(i\phi)+(\Sigma_fe_f\tilde{J}_f^{x})\cdot\Sigma_{i=1}^3a_{gh,x}^i\text{sin}(i\phi)\big)I_{gh}^*\Big]\Big\},
\end{align}

the harmonic coefficients in Eq.~\eqref{priQCDh} read:
\begin{align*}
a_{qh1,y}^2=&\frac{32 e_q Q \mathbf{q_\perp^2} \sqrt{1-y} (y-2)}{y^2},\nonumber \\
a_{qh1,y}^1=&-\frac{32 Q^2 \big|q_\perp\big| (y-2) z \bar{z}\left(\mu ^2+\mathbf{q_\perp^2}\right)}{y\left(m_q^2+\mathbf{q_\perp^2}\right)},\nonumber \\
a_{qh1,y}^0=&\frac{32 e_q \mu ^2 Q \sqrt{1-y} (y-2) }{y^2},\nonumber \\
a_{qh1,x}^2=&-a_{qh1,y}^2,\nonumber \\
a_{qh1,x}^1=&\frac{32 Q^2 \big|q_\perp\big| (y-2) z \bar{z}\left(m_q^2-Q^2 z \bar{z}+\mathbf{q_\perp^2}\right)}{y \left(m_q^2+\mathbf{q_\perp^2}\right)},\nonumber \\
a_{qh2,y}^3=&-\frac{32 e_q \big|q_\perp\big|\mathbf{q_\perp^2} (y-1) }{y^2}\nonumber \\
a_{qh2,y}^2=&\frac{32 Q \mathbf{q_\perp^2} \sqrt{1-y}\left(e_q (2-y)\bar{z} \left(m_q^2+\mathbf{q_\perp^2}\right)-y z \bar{z} \left(\mu ^2+\mathbf{q_\perp^2}\right)\right)}{y^2 \left(m_q^2+\mathbf{q_\perp^2}\right)},\nonumber \\
a_{qh2,y}^1=&\frac{16 e_q\big|q_\perp\big| \left(\mu^2 (y (y\bar{z}-2)+2)+\bar{z}\mathbf{q_\perp^2} ((y-2) y+2)\right)}{y^2 z},\nonumber \\
a_{qh2,y}^0=&-\frac{16 Q \mathbf{q_\perp^2} \left(y ((y-2) y+2) \bar{z}^2 \left(\mu ^2+\mathbf{q_\perp^2}\right)+2 e_q(y-2) (y-1) \bar{z}\left(m_q^2+\mathbf{q_\perp^2}\right)\right)}{\sqrt{1-y} y^2 \left(m_q^2+\mathbf{q_\perp^2}\right)},
\end{align*}
\begin{align} \label{priQCDhcoeff}
\mathmakebox[2.0cm][r]a_{qh2,x}^3=&-a_{qh2,y}^3,\nonumber \\
a_{qh2,x}^2=&-a_{qh2,y}^2-\frac{64 Q^3 \mathbf{q_\perp^2} \sqrt{1-y} z^2 \bar{z}^2}{y \left(m_q^2+\mathbf{q_\perp^2}\right)},\nonumber \\
a_{qh2,x}^1=&-a_{qh2,y}^1-\frac{64 \big|q_\perp\big|\left(e_q \mu ^2 (y-1)+Q^2 (y-2) y \bar{z}^2 z\right)}{y^2},\nonumber \\
a_{qh3,y}^3=&a_{qh2,y}^3,\nonumber \\
a_{qh3,y}^2=&a_{qh2,y}^2+\frac{32 e_q Q \mathbf{q_\perp^2} \sqrt{1-y} (y-2)}{y^2},\nonumber \\
a_{qh3,y}^1=&a_{qh2,y}^1+\frac{16 e_q \big|q_\perp\big| ((y-2) y+2) (2 z-1)\left(\mu ^2+\mathbf{q_\perp^2}\right) }{y^2  z\bar{z}},\nonumber \\
a_{qh3,y}^0=&-\frac{16 Q \mathbf{q_\perp^2} z \left(y ((y-2) y+2) z \left(\mu ^2+\mathbf{q_\perp^2}\right)-2 e_q (y-2) (y-1) \left(m_q^2+\mathbf{q_\perp^2}\right)\right)}{\sqrt{1-y} y^2 \left(m_q^2+\mathbf{q_\perp^2}\right)},\nonumber \\
a_{qh3,x}^3=&-a_{qh2,y}^3,\nonumber \\
a_{qh3,x}^2=&-a_{qh3,y}^2-\frac{64 Q^3 \mathbf{q_\perp^2} \sqrt{1-y} z^2 \bar{z}^2 }{y\left(m_q^2+\mathbf{q_\perp^2}\right)},\nonumber \\
a_{qh3,x}^1=&-\frac{16 \big|q_\perp\big|  \left(\text{eq} \left(\mu^2 \left((y-2)^2 z+2 (y-1)\right)+\mathbf{q_\perp^2} ((y-2) y+2) z\right)-4 Q^2 (y-2) y z^2 \bar{z}^2\right)}{y^2 \bar{z}},\nonumber \\
a_{gh,y}^3=&-\frac{32 e_q \big|q_\perp\big|\mathbf{q_\perp^2}(y-1)}{y^2},\nonumber \\
a_{gh,y}^2=&\frac{16 Q \mathbf{q_\perp^2} \sqrt{1-y} \left(e_q (y-2) (2 z-1) \left(m_q^2+\mathbf{q_\perp^2}\right)-2 y z \bar{z} \left(\mu ^2+\mathbf{q_\perp^2}\right)\right)}{y^2 \left(m_q^2+\mathbf{q_\perp^2}\right)},\nonumber \\
a_{gh,y}^1=&-\frac{4 e_q \big|q_\perp\big|\left((-y^2 (1-2 z)^2-(y-2)^2) \mu^2+2 \mathbf{q_\perp^2} ((y-2) y+2) (2 z \bar{z}-1)\right)}{ y^2 z \bar{z}},\nonumber \\
a_{gh,y}^0=&\frac{8 Q\mathbf{q_\perp^2} }{\sqrt{1-y} y^2 \left(m_q^2+\mathbf{q_\perp^2}\right)}\Big[2 e_q(y-2) (y-1) (2 z-1) \left(m_q^2+\mathbf{q_\perp^2}\right)+y ((y-2) y+2) \left(\mu ^2+\mathbf{q_\perp^2}\right) (2 z \bar{z}-1)\Big],\nonumber \\
a_{gh,x}^3=&-a_{gh,y}^3,\nonumber \\
a_{gh,x}^2=&-a_{gh,y}^2-\frac{64 Q^3 \mathbf{q_\perp^2} \sqrt{1-y} z^2\bar{z}^2 }{y \left(m_q^2+\mathbf{q_\perp^2}\right)},\nonumber \\
a_{gh,x}^1=&-\frac{4  \big|q_\perp\big| }{y^2 z \bar{z}}\Big[e_q \left(\mu^2(y-2)^2 (1-2 z)^2-2 \mathbf{q_\perp^2} ((y-2) y+2) (2 z \bar{z}-1)+\mu ^2 y^2\right)\nonumber \\
 &-8 Q^2 (y-2) y z^2 (2 z-1) \bar{z}^2\Big],
\end{align}
$m_q$'s in Eq.~\eqref{priQCDu},Eq.~\eqref{priQCDucoeff},Eq.~\eqref{priQCDh},Eq.~\eqref{priQCDhcoeff} are only nonzero when the two conditions are satisfied:

(1)They appear in the coefficient of the interference between an EFF and a gluon CFF.

(2)When one deals with the charm or bottom dijet production.

\subsection{\texorpdfstring{$\big|M^{q\bar{q}}_{\text{QED}}\big|^2$}{Pure QED}} \label{pureQED_form}
We have:
\begin{align}
\big|M_{\text{QED}}^{q\bar{q}}\big|^2=&\frac{1}{2}\cdot\frac{256\pi^4\alpha_{\text{em}}^4e_q^2N_c}{t^2(\mu^2+\mathbf{q_\perp^2})^4}\Big[(\Sigma_fe_f\tilde{J}_f^x)(\Sigma_{f^\prime}e_{f^\prime}\tilde{J}_{f^\prime}^{x*})\cdot\Sigma_{i=0}^4a_{xx}^i\text{cos}(i\phi)+(\Sigma_fe_f\tilde{J}_f^y)(\Sigma_{f^\prime}e_{f^\prime}\tilde{J}_{f^\prime}^{y*})\cdot\Sigma_{i=0}^4a_{yy}^i\text{cos}(i\phi)\nonumber \\
&+\text{Re}\Big[(\Sigma_fe_f\tilde{J}_f^x)(\Sigma_{f^\prime}e_{f^\prime}\tilde{J}_{f^\prime}^{y*})\Big]\cdot\Sigma_{i=1}^4a_{xy}^i\text{sin}(i\phi)\Big],
\end{align}
the harmonic coefficients read:
\begin{align*}
a_{xx}^0=&\frac{1}{Q^2 (y-1) y^2 z \bar{z} \left(m_q^2+\mathbf{q_\perp^2}\right)^2}8\left(\mu ^2+\mathbf{q_\perp^2}\right)^2\Bigg\{e_q^2(y-1)(m_q^2+\mathbf{q_\perp^2})\Big[(m_q^4+2m_q^2Q^2z\bar{z}+Q^4z^2\bar{z}^2) \nonumber \\
&\cdot\left((y-1)^2+1-2(y-2)^2z\bar{z}\right)+(2 m_q^2 \mathbf{q_\perp^2}+\mathbf{q_\perp^4}(1-2 z \bar{z})) ((y-2) y+2)-16 Q^2 \mathbf{q_\perp^2} (y-1) z^2 \bar{z}^2\Big]\nonumber \\
&-4e_qQ^2(y-2)(y-1)yz^2(2 z-1)\bar{z}^2(m_q^2+\mathbf{q_\perp^2})\Big[2 m_q^4+m_q^2 \left(2Q^2z\bar{z}+\mathbf{q_\perp^2}\right)-\mathbf{q_\perp^2}(\mathbf{q_\perp^2}-3Q^2z\bar{z})\Big]\nonumber \\
&-Q^2y^2z^2\bar{z}^2\Big[(m_q^6+m_q^4\mathbf{q_\perp^2}(3-2z\bar{z})+m_q^2Q^4z^2\bar{z}^2+m_q^2\mathbf{q_\perp^4}(3-4z\bar{z})+Q^4\mathbf{q_\perp^2}z^2\bar{z}^2(1-2z\bar{z})\nonumber \\
&+\mathbf{q_\perp^6}(1-2z\bar{z})) ((y-2) y+2)-2m_q^4 Q^2 z \bar{z} \left(y^2-2 (y-1) (1-8 z \bar{z})\right)-4m_q^2Q^2 \mathbf{q_\perp^2} z \bar{z} \nonumber \\
&\cdot\left(y^2 (1-z \bar{z})-2 (y-1) (1-9 z \bar{z})\right)-2 Q^2 \mathbf{q_\perp^4} z \bar{z} \left(y^2 (1-2 z \bar{z})+(y-1) (20 z \bar{z}-2)\right)\Big]\Bigg\},\nonumber \\
a_{xx}^1=&\frac{1}{Q \sqrt{1-y} y^2 \left(m_q^2+\mathbf{q_\perp^2}\right)}16\big|q_\perp\big|\left(\mu ^2+\mathbf{q_\perp^2}\right)^2\Bigg\{2 e_q^2 (1-y) (2-y) (1-2 z) \left(m_q^2+\mathbf{q_\perp^2}\right) (2 m_q^2+2 Q^2 z \bar{z}\nonumber \\
&-\mathbf{q_\perp^2})-e_q y \left(\mu ^2+\mathbf{q_\perp^2}\right)^2\big(y(-2yz\bar{z}+y-2)+2\big)-4 Q^2 (2-y) y^2 (1-2 z) z^2 \bar{z}^2 \left(m_q^2-Q^2 z \bar{z}+\mathbf{q_\perp^2}\right)\nonumber \\
&-4e_qyz\bar{z}\Big[m_q^4 ((y-2)y+2)+m_q^2\left(Q^2\left(((y-8)y+8)z\bar{z}-(y-1)^2-1\right)+\mathbf{q_\perp^2} (y(y-1)+1)\right)\nonumber \\
&+2Q^4 (y-1) z^2 \bar{z}^2-Q^2 \mathbf{q_\perp^2} \left(y^2 (1-2 z \bar{z})+(y-1)(11 z \bar{z}-2)\right)+\mathbf{q_\perp^4} (y-1)\Big]\Bigg\},\nonumber \\
a_{xx}^2=&\frac{16 \mathbf{q_\perp^2}}{Q^2 y^2 \left(m_q^2+\mathbf{q_\perp^2}\right)^2}\Bigg\{e_q^2 \left(m_q^2+\mathbf{q_\perp^2}\right)^2 \left(\mu ^2+\mathbf{q_\perp^2}\right)^2\Big[2 m_q^2 (y-2)^2+Q^2(-(y-1)^2-1+2((y-8)y+8)z\bar{z})\Big]\nonumber \\
&-2 e_q Q^2 (y-2) y z (2 z-1) \bar{z} \left(m_q^2+\mathbf{q_\perp^2}\right) \left(\mu ^2+\mathbf{q_\perp^2}\right)^2 \left(Q^2 z \bar{z}-3 \left(m_q^2+\mathbf{q_\perp^2}\right)\right)-2 Q^2 y^2 z^2 \bar{z}^2 \nonumber \\
&\cdot\left((m_q^2+\mathbf{q_\perp^2}\right)^2-Q^4 z^2 \bar{z}^2)^2\Bigg\},\nonumber \\
a_{xx}^3=&-\frac{32\mathbf{q_\perp^2}\big|q_\perp\big| \sqrt{1-y} \left(\mu ^2+\mathbf{q_\perp^2}\right)^2 \Big[e_q^2 (y-2) (2 z-1) \left(m_q^2+\mathbf{q_\perp^2}\right)-2e_q y z \bar{z} \left(m_q^2-Q^2 z \bar{z}+\mathbf{q_\perp^2}\right)\Big]}{Q y^2 \left(m_q^2+\mathbf{q_\perp^2}\right)},\nonumber \\
a_{xx}^4=&\frac{32e_q^2 \mathbf{q_\perp^4} (y-1)  \left(\mu ^2+\mathbf{q_\perp^2}\right)^2}{Q^2 y^2},\nonumber \\
a_{yy}^0=&\frac{8\left(\mu ^2+\mathbf{q_\perp^2}\right)^2}{Q^2(y-1)y^2z\bar{z}(m_q^2+\mathbf{q_\perp^2})^2}\Bigg\{e_q^2(y-1)\left(m_q^2+\mathbf{q_\perp^2}\right)^2\Big[2\mathbf{q_\perp^2}\left(2 m_q^2 y^2 z \bar{z}-Q^2z\bar{z} \left((y-1)^2+1-2 (y-2)^2 z\bar{z}\right)\right)\nonumber \\
&-\mathbf{q_\perp^4} ((y-2) y+2) (2 z \bar{z}-1)+\mu^2(\mu^2+2\mathbf{q_\perp^2})\big(y (-2 y z \bar{z}+y-2)+2\big)\Big]+4 e_q Q^2 \mathbf{q_\perp^2} (y-2) (y-1) y(2 z-1)\nonumber \\
&\cdot z^2 \bar{z}^2 \left(m_q^2+\mathbf{q_\perp^2}\right) \left(\mu ^2+\mathbf{q_\perp^2}\right)-Q^2 y^2 ((y-2) y+2) z^2 \bar{z}^2 \left(\mu ^2+\mathbf{q_\perp^2}\right)^2 \left(m_q^2+\mathbf{q_\perp^2} (1-2 z \bar{z})\right)\Bigg\},\nonumber \\
a_{yy}^1=&\frac{16e_q\big|q_\perp\big|\left(\mu ^2+\mathbf{q_\perp^2}\right)^2}{Q \sqrt{1-y} y^2 \left(m_q^2+\mathbf{q_\perp^2}\right)}\Bigg\{-y(\mu ^2+\mathbf{q_\perp^2})^2(y (-2 y z \bar{z}+y-2)+2) +2e_q\mathbf{q_\perp^2} (y-2) (y-1) (2 z-1) \left(m_q^2+\mathbf{q_\perp^2}\right)\nonumber \\
&-4 \mathbf{q_\perp^2} (y-1) y z \bar{z} \left(\mu ^2+\mathbf{q_\perp^2}\right)\Bigg\},\nonumber \\
a_{yy}^2=&\frac{16 \mathbf{q_\perp^2} \left(\mu ^2+\mathbf{q_\perp^2}\right)^2}{Q^2 y^2 \left(m_q^2+\mathbf{q_\perp^2}\right)^2}\Bigg\{e_q^2 \left(m_q^2+\mathbf{q_\perp^2}\right)^2 \Big[Q^2 \left((y-1)^2+1-2(y-2)^2z\bar{z}\right)-2 m_q^2 y^2\Big]-2 e_q Q^2 (y-2) y z (2 z-1) \bar{z} \nonumber \\
&\cdot\left(m_q^2+\mathbf{q_\perp^2}\right)\left(\mu ^2+\mathbf{q_\perp^2}\right)+2 Q^2 y^2 z^2 \bar{z}^2 \left(\mu ^2+\mathbf{q_\perp^2}\right)^2\Bigg\},
\end{align*}
\begin{align} \label{QEDcoeff}
a_{yy}^3=&-a_{xx}^3-\frac{128 e_qQ \mathbf{q_\perp^2}\big|q_\perp\big| \sqrt{1-y} z^2 \bar{z}^2 \left(\mu ^2+\mathbf{q_\perp^2}\right)^2}{y \left(m_q^2+\mathbf{q_\perp^2}\right)},\nonumber \\
a_{yy}^4=&-a_{xx}^4,\nonumber \\
a_{xy}^1=&\frac{16 \big|q_\perp\big| \left(\mu ^2+\mathbf{q_\perp^2}\right)^2}{Q \sqrt{1-y} y^2 \left(m_q^2+\mathbf{q_\perp^2}\right)}\Bigg\{-4 e_q^2 \mu ^2 (y-2) (y-1) (2 z-1) \left(m_q^2+\mathbf{q_\perp^2}\right)-4e_qyz\bar{z}\Big[m_q^4 (y (y -2)+2)\nonumber \\
&-m_q^2 \left(-Q^2(y(y-12)+12)z\bar{z}+((y-2)y+2)(Q^2-\mathbf{q_\perp^2})\right)-2 Q^4 (y-1) z^2 \bar{z}^2\nonumber \\
&-Q^2 \mathbf{q_\perp^2} \left((y-1)^2+1-2 ((y-6) y+6) z\bar{z}\right)\Big]-4 Q^2 (y-2) y^2 z^2 (2 z-1) \bar{z}^2 \left(\mu ^2+\mathbf{q_\perp^2}\right)\Bigg\}\nonumber \\
a_{xy}^2=&\frac{32\mathbf{q_\perp^2}\left(\mu ^2+\mathbf{q_\perp^2}\right)^2}{Q^2y^2(m_q^2+\mathbf{q_\perp^2})^2}\Bigg\{e_q^2 \left(m_q^2+\mathbf{q_\perp^2}\right)^2 \Big[((y-2) y+2) \left(2 m_q^2-Q^2\right)+2 Q^2 ((y-6) y+6) z\bar{z}\Big]+4 e_qQ^2 (y-2) y \nonumber \\
&\cdot(2 z-1) z\bar{z}\left(m_q^2+\mathbf{q_\perp^2}\right)^2-2 Q^2 y^2 z^2 \bar{z}^2 \left(\left(m_q^2+\mathbf{q_\perp^2}\right)^2-Q^4 z^2 \bar{z}^2\right)\Bigg\},\nonumber \\
a_{xy}^3=&2a_{xx}^3+\frac{128 e_qQ \mathbf{q_\perp^2}\big|q_\perp\big| \sqrt{1-y} z^2 \bar{z}^2 \left(\mu ^2+\mathbf{q_\perp^2}\right)^2}{y \left(m_q^2+\mathbf{q_\perp^2}\right)},\nonumber \\
a_{xy}^4=&2a_{xx}^4,
\end{align}
the $m_q$'s in Eq.~\eqref{QEDcoeff} are only nonzero for charm or bottom dijet production.
\bibliography{bibliography}

@article{Braun:2005rg,
    author = "Braun, V. M. and Ivanov, D. Yu.",
    title = "{Exclusive diffractive electroproduction of dijets in collinear factorization}",
    eprint = "hep-ph/0505263",
    archivePrefix = "arXiv",
    doi = "10.1103/PhysRevD.72.034016",
    journal = "Phys. Rev. D",
    volume = "72",
    pages = "034016",
    year = "2005"
}

@article{Diehl:2003ny,
    author = "Diehl, M.",
    title = "{Generalized parton distributions}",
    eprint = "hep-ph/0307382",
    archivePrefix = "arXiv",
    reportNumber = "DESY-THESIS-2003-018",
    doi = "10.1016/j.physrep.2003.08.002",
    journal = "Phys. Rept.",
    volume = "388",
    pages = "41--277",
    year = "2003"
}

@article{Pedrak:2020mfm,
    author = "Pedrak, A. and Pire, B. and Szymanowski, L. and Wagner, J.",
    title = "{Electroproduction of a large invariant mass photon pair}",
    eprint = "2003.03263",
    archivePrefix = "arXiv",
    primaryClass = "hep-ph",
    reportNumber = "CPHT-RR010.022020",
    doi = "10.1103/PhysRevD.101.114027",
    journal = "Phys. Rev. D",
    volume = "101",
    number = "11",
    pages = "114027",
    year = "2020"
}

@article{Pedrak:2017cpp,
    author = "Pedrak, A. and Pire, B. and Szymanowski, L. and Wagner, J.",
    title = "{Hard photoproduction of a diphoton with a large invariant mass}",
    eprint = "1708.01043",
    archivePrefix = "arXiv",
    primaryClass = "hep-ph",
    reportNumber = "CPHT-RR-047-082017",
    doi = "10.1103/PhysRevD.96.074008",
    journal = "Phys. Rev. D",
    volume = "96",
    number = "7",
    pages = "074008",
    year = "2017",
    note = "[Erratum: Phys.Rev.D 100, 039901 (2019)]"
}

@article{Grocholski:2021man,
    author = "Grocholski, Oskar and Pire, Bernard and Sznajder, Pawe{\l} and Szymanowski, Lech and Wagner, Jakub",
    title = "{Collinear factorization of diphoton photoproduction at next to leading order}",
    eprint = "2110.00048",
    archivePrefix = "arXiv",
    primaryClass = "hep-ph",
    reportNumber = "CPHT-RR077.092021",
    doi = "10.1103/PhysRevD.104.114006",
    journal = "Phys. Rev. D",
    volume = "104",
    number = "11",
    pages = "114006",
    year = "2021"
}

@article{Diehl:2001pm,
    author = "Diehl, M.",
    title = "{Generalized parton distributions with helicity flip}",
    eprint = "hep-ph/0101335",
    archivePrefix = "arXiv",
    reportNumber = "DESY-01-009",
    doi = "10.1007/s100520100635",
    journal = "Eur. Phys. J. C",
    volume = "19",
    pages = "485--492",
    year = "2001"
}

@article{Grammer:1973db,
    author = "Grammer, Jr., G. and Yennie, D. R.",
    title = "{Improved treatment for the infrared divergence problem in quantum electrodynamics}",
    doi = "10.1103/PhysRevD.8.4332",
    journal = "Phys. Rev. D",
    volume = "8",
    pages = "4332--4344",
    year = "1973"
}

@article{Mueller:2005ed,
    author = "Mueller, Dieter and Schafer, A.",
    title = "{Complex conformal spin partial wave expansion of generalized parton distributions and distribution amplitudes}",
    eprint = "hep-ph/0509204",
    archivePrefix = "arXiv",
    doi = "10.1016/j.nuclphysb.2006.01.019",
    journal = "Nucl. Phys. B",
    volume = "739",
    pages = "1--59",
    year = "2006"
}

@article{Mankiewicz:1999tt,
    author = "Mankiewicz, L. and Piller, G.",
    title = "{Comments on exclusive electroproduction of transversely polarized vector mesons}",
    eprint = "hep-ph/9905287",
    archivePrefix = "arXiv",
    reportNumber = "TUM-T39-99-6",
    doi = "10.1103/PhysRevD.61.074013",
    journal = "Phys. Rev. D",
    volume = "61",
    pages = "074013",
    year = "2000"
}

@article{Radyushkin:1998es,
    author = "Radyushkin, A. V.",
    title = "{Double distributions and evolution equations}",
    eprint = "hep-ph/9805342",
    archivePrefix = "arXiv",
    reportNumber = "JLAB-THY-98-16",
    doi = "10.1103/PhysRevD.59.014030",
    journal = "Phys. Rev. D",
    volume = "59",
    pages = "014030",
    year = "1999"
}

@article{Belitsky:2005qn,
    author = "Belitsky, A. V. and Radyushkin, A. V.",
    title = "{Unraveling hadron structure with generalized parton distributions}",
    eprint = "hep-ph/0504030",
    archivePrefix = "arXiv",
    reportNumber = "JLAB-THY-04-34",
    doi = "10.1016/j.physrep.2005.06.002",
    journal = "Phys. Rept.",
    volume = "418",
    pages = "1--387",
    year = "2005"
}

@article{Ji:1996nm,
    author = "Ji, Xiang-Dong",
    title = "{Deeply virtual Compton scattering}",
    eprint = "hep-ph/9609381",
    archivePrefix = "arXiv",
    reportNumber = "UMD-PP-97-26, MIT-CTP-2568",
    doi = "10.1103/PhysRevD.55.7114",
    journal = "Phys. Rev. D",
    volume = "55",
    pages = "7114--7125",
    year = "1997"
}

@article{Guo:2025muf,
    author = "Guo, Yuxun and Aslan, Fatma P. and Ji, Xiangdong and Santiago, M. Gabriel",
    title = "{First Global Extraction of Generalized Parton Distributions from Experiment and Lattice Data with Next-to-Leading-Order Accuracy}",
    eprint = "2509.08037",
    archivePrefix = "arXiv",
    primaryClass = "hep-ph",
    doi = "10.1103/qct5-y7rp",
    journal = "Phys. Rev. Lett.",
    volume = "135",
    number = "26",
    pages = "261903",
    year = "2025"
}

@article{Muller:1994ses,
    author = {M{\"u}ller, Dieter and Robaschik, D. and Geyer, B. and Dittes, F. -M. and Ho{\v{r}}ej{\v{s}}i, J.},
    title = "{Wave functions, evolution equations and evolution kernels from light ray operators of QCD}",
    eprint = "hep-ph/9812448",
    archivePrefix = "arXiv",
    reportNumber = "NTZ-6-91, NTZ-91-6",
    doi = "10.1002/prop.2190420202",
    journal = "Fortsch. Phys.",
    volume = "42",
    pages = "101--141",
    year = "1994"
}

@article{Radyushkin:1997ki,
    author = "Radyushkin, A. V.",
    title = "{Nonforward parton distributions}",
    eprint = "hep-ph/9704207",
    archivePrefix = "arXiv",
    reportNumber = "JLAB-THY-97-10",
    doi = "10.1103/PhysRevD.56.5524",
    journal = "Phys. Rev. D",
    volume = "56",
    pages = "5524--5557",
    year = "1997"
}

@article{Ji:1996ek,
    author = "Ji, Xiang-Dong",
    title = "{Gauge-Invariant Decomposition of Nucleon Spin}",
    eprint = "hep-ph/9603249",
    archivePrefix = "arXiv",
    reportNumber = "MIT-CTP-2517",
    doi = "10.1103/PhysRevLett.78.610",
    journal = "Phys. Rev. Lett.",
    volume = "78",
    pages = "610--613",
    year = "1997"
}

@article{Polyakov:2002yz,
    author = "Polyakov, M. V.",
    title = "{Generalized parton distributions and strong forces inside nucleons and nuclei}",
    eprint = "hep-ph/0210165",
    archivePrefix = "arXiv",
    reportNumber = "RUB-TP2-14-02",
    doi = "10.1016/S0370-2693(03)00036-4",
    journal = "Phys. Lett. B",
    volume = "555",
    pages = "57--62",
    year = "2003"
}

@article{Ji:1994av,
    author = "Ji, Xiang-Dong",
    title = "{A QCD analysis of the mass structure of the nucleon}",
    eprint = "hep-ph/9410274",
    archivePrefix = "arXiv",
    reportNumber = "MIT-CTP-2368",
    doi = "10.1103/PhysRevLett.74.1071",
    journal = "Phys. Rev. Lett.",
    volume = "74",
    pages = "1071--1074",
    year = "1995"
}

@article{Collins:1996fb,
    author = "Collins, John C. and Frankfurt, Leonid and Strikman, Mark",
    title = "{Factorization for hard exclusive electroproduction of mesons in QCD}",
    eprint = "hep-ph/9611433",
    archivePrefix = "arXiv",
    reportNumber = "CERN-TH-96-314, PSU-TH-168",
    doi = "10.1103/PhysRevD.56.2982",
    journal = "Phys. Rev. D",
    volume = "56",
    pages = "2982--3006",
    year = "1997"
}

@article{Ji:2013dva,
    author = "Ji, Xiangdong",
    title = "{Parton Physics on a Euclidean Lattice}",
    eprint = "1305.1539",
    archivePrefix = "arXiv",
    primaryClass = "hep-ph",
    doi = "10.1103/PhysRevLett.110.262002",
    journal = "Phys. Rev. Lett.",
    volume = "110",
    pages = "262002",
    year = "2013"
}

@article{Ji:2014gla,
    author = "Ji, Xiangdong",
    title = "{Parton Physics from Large-Momentum Effective Field Theory}",
    eprint = "1404.6680",
    archivePrefix = "arXiv",
    primaryClass = "hep-ph",
    doi = "10.1007/s11433-014-5492-3",
    journal = "Sci. China Phys. Mech. Astron.",
    volume = "57",
    pages = "1407--1412",
    year = "2014"
}

@article{Ji:2020ect,
    author = "Ji, Xiangdong and Liu, Yu-Sheng and Liu, Yizhuang and Zhang, Jian-Hui and Zhao, Yong",
    title = "{Large-momentum effective theory}",
    eprint = "2004.03543",
    archivePrefix = "arXiv",
    primaryClass = "hep-ph",
    doi = "10.1103/RevModPhys.93.035005",
    journal = "Rev. Mod. Phys.",
    volume = "93",
    number = "3",
    pages = "035005",
    year = "2021"
}

@article{Radyushkin:2017cyf,
    author = "Radyushkin, A. V.",
    title = "{Quasi-parton distribution functions, momentum distributions, and pseudo-parton distribution functions}",
    eprint = "1705.01488",
    archivePrefix = "arXiv",
    primaryClass = "hep-ph",
    reportNumber = "JLAB-THY-17-2455",
    doi = "10.1103/PhysRevD.96.034025",
    journal = "Phys. Rev. D",
    volume = "96",
    number = "3",
    pages = "034025",
    year = "2017"
}

@article{Anderle:2021wcy,
    author = "Anderle, Daniele P. and others",
    title = "{Electron-ion collider in China}",
    eprint = "2102.09222",
    archivePrefix = "arXiv",
    primaryClass = "nucl-ex",
    reportNumber = "Frontiers of Physics, Volume 16 Issue (6):64701, 2021",
    doi = "10.1007/s11467-021-1062-0",
    journal = "Front. Phys. (Beijing)",
    volume = "16",
    number = "6",
    pages = "64701",
    year = "2021"
}

@article{AbdulKhalek:2021gbh,
    author = "Abdul Khalek, R. and others",
    title = "{Science Requirements and Detector Concepts for the Electron-Ion Collider}: {EIC Yellow Report}",
    eprint = "2103.05419",
    archivePrefix = "arXiv",
    primaryClass = "physics.ins-det",
    reportNumber = "BNL-220990-2021-FORE, JLAB-PHY-21-3198, LA-UR-21-20953",
    doi = "10.1016/j.nuclphysa.2022.122447",
    journal = "Nucl. Phys. A",
    volume = "1026",
    pages = "122447",
    year = "2022"
}

@article{Accardi:2012qut,
    author = "Accardi, A. and others",
    editor = "Deshpande, A. and Meziani, Z. E. and Qiu, J. W.",
    title = "{Electron Ion Collider: The Next QCD Frontier}: {Understanding the glue that binds us all}",
    eprint = "1212.1701",
    archivePrefix = "arXiv",
    primaryClass = "nucl-ex",
    reportNumber = "BNL-98815-2012-JA, JLAB-PHY-12-1652",
    doi = "10.1140/epja/i2016-16268-9",
    journal = "Eur. Phys. J. A",
    volume = "52",
    number = "9",
    pages = "268",
    year = "2016"
}

@article{Collins:1998be,
    author = "Collins, John C. and Freund, Andreas",
    title = "{Proof of factorization for deeply virtual Compton scattering in QCD}",
    eprint = "hep-ph/9801262",
    archivePrefix = "arXiv",
    reportNumber = "PSU-TH-192",
    doi = "10.1103/PhysRevD.59.074009",
    journal = "Phys. Rev. D",
    volume = "59",
    pages = "074009",
    year = "1999"
}

@article{Chu:2025kew,
    author = "Chu, Min-Huan and Cola{\c{c}}o, Manuel and Bhattacharya, Shohini and Cichy, Krzysztof and Constantinou, Martha and Metz, Andreas and Steffens, Fernanda",
    title = "{Generalized parton distributions from lattice QCD with asymmetric momentum transfer: Unpolarized quarks at nonzero skewness}",
    eprint = "2508.17998",
    archivePrefix = "arXiv",
    primaryClass = "hep-lat",
    doi = "10.1103/ts5s-hvb1",
    journal = "Phys. Rev. D",
    volume = "112",
    number = "9",
    pages = "094510",
    year = "2025"
}

@article{Bhattacharya:2022aob,
    author = "Bhattacharya, Shohini and Cichy, Krzysztof and Constantinou, Martha and Dodson, Jack and Gao, Xiang and Metz, Andreas and Mukherjee, Swagato and Scapellato, Aurora and Steffens, Fernanda and Zhao, Yong",
    title = "{Generalized parton distributions from lattice QCD with asymmetric momentum transfer: Unpolarized quarks}",
    eprint = "2209.05373",
    archivePrefix = "arXiv",
    primaryClass = "hep-lat",
    doi = "10.1103/PhysRevD.106.114512",
    journal = "Phys. Rev. D",
    volume = "106",
    number = "11",
    pages = "114512",
    year = "2022"
}

@article{Lin:2020rxa,
    author = "Lin, Huey-Wen",
    title = "{Nucleon Tomography and Generalized Parton Distribution at Physical Pion Mass from Lattice QCD}",
    eprint = "2008.12474",
    archivePrefix = "arXiv",
    primaryClass = "hep-ph",
    reportNumber = "MSUHEP-20-014, MSUHEP-20-014",
    doi = "10.1103/PhysRevLett.127.182001",
    journal = "Phys. Rev. Lett.",
    volume = "127",
    number = "18",
    pages = "182001",
    year = "2021"
}

@article{Alexandrou:2020zbe,
    author = "Alexandrou, Constantia and Cichy, Krzysztof and Constantinou, Martha and Hadjiyiannakou, Kyriakos and Jansen, Karl and Scapellato, Aurora and Steffens, Fernanda",
    title = "{Unpolarized and helicity generalized parton distributions of the proton within lattice QCD}",
    eprint = "2008.10573",
    archivePrefix = "arXiv",
    primaryClass = "hep-lat",
    reportNumber = "DESY-20-150",
    doi = "10.1103/PhysRevLett.125.262001",
    journal = "Phys. Rev. Lett.",
    volume = "125",
    number = "26",
    pages = "262001",
    year = "2020"
}

@article{Chu:2025jsi,
    author = "Chu, Min-Huan and Cichy, Krzysztof and Constantinou, Martha and Sznajder, Pawe{\l} and Wagner, Jakub",
    title = "{A unified neural-network framework for nucleon imaging from numerical simulations of QCD}",
    eprint = "2509.15931",
    archivePrefix = "arXiv",
    primaryClass = "hep-lat",
    doi = "10.1007/JHEP05(2026)210",
    journal = "JHEP",
    volume = "05",
    pages = "210",
    year = "2026"
}

@article{ZEUS:2015sns,
    author = "Abramowicz, H. and others",
    collaboration = "ZEUS",
    title = "{Production of exclusive dijets in diffractive deep inelastic scattering at HERA}",
    eprint = "1505.05783",
    archivePrefix = "arXiv",
    primaryClass = "hep-ex",
    reportNumber = "DESY-15-070",
    doi = "10.1140/epjc/s10052-015-3849-z",
    journal = "Eur. Phys. J. C",
    volume = "76",
    number = "1",
    pages = "16",
    year = "2016"
}

@article{Bartels:1996ne,
    author = {Bartels, Jochen and Lotter, H. and W{\"u}sthoff, M.},
    title = "{Quark-antiquark production in DIS diffractive dissociation}",
    eprint = "hep-ph/9602363",
    archivePrefix = "arXiv",
    reportNumber = "DESY-96-026, ANL-HEP-PR-96-11",
    doi = "10.1016/0370-2693(96)00412-1",
    journal = "Phys. Lett. B",
    volume = "379",
    pages = "239--248",
    year = "1996",
    note = "[Erratum: Phys.Lett.B 382, 449--449 (1996)]"
}

@article{Bartels:1996tc,
    author = "Bartels, Jochen and Ewerz, C. and Lotter, H. and Wusthoff, M.",
    title = "{Azimuthal distribution of quark - anti-quark jets in DIS diffractive dissociation}",
    eprint = "hep-ph/9605356",
    archivePrefix = "arXiv",
    reportNumber = "DESY-96-085, ANL-HEP-PR-96-39",
    doi = "10.1016/0370-2693(96)81071-9",
    journal = "Phys. Lett. B",
    volume = "386",
    pages = "389--396",
    year = "1996"
}

@article{Bartels:1999tn,
    author = "Bartels, Jochen and Jung, H. and Wusthoff, M.",
    title = "{Quark - anti-quark gluon jets in DIS diffractive dissociation}",
    eprint = "hep-ph/9903265",
    archivePrefix = "arXiv",
    reportNumber = "DESY-99-027, DTP-99-10, LUNFD6-NFFL-7166-1999",
    doi = "10.1007/s100520050618",
    journal = "Eur. Phys. J. C",
    volume = "11",
    pages = "111--125",
    year = "1999"
}

@article{Boussarie:2019ero,
    author = "Boussarie, R. and Grabovsky, A. V. and Szymanowski, L. and Wallon, S.",
    title = "{Towards a complete next-to-logarithmic description of forward exclusive diffractive dijet electroproduction at HERA: real corrections}",
    eprint = "1905.07371",
    archivePrefix = "arXiv",
    primaryClass = "hep-ph",
    reportNumber = "LPT-Orsay-19-22",
    doi = "10.1103/PhysRevD.100.074020",
    journal = "Phys. Rev. D",
    volume = "100",
    number = "7",
    pages = "074020",
    year = "2019"
}

@article{Nikolaev:1994cd,
    author = "Nikolaev, Nikolai N. and Zakharov, B. G.",
    title = "{Splitting the pomeron into two jets: A Novel process at HERA}",
    eprint = "hep-ph/9403281",
    archivePrefix = "arXiv",
    reportNumber = "KFA-IKP-TH-1994-16",
    doi = "10.1016/0370-2693(94)90876-1",
    journal = "Phys. Lett. B",
    volume = "332",
    pages = "177--183",
    year = "1994"
}

@article{Linek:2024dzs,
    author = {Linek, Barbara and {\L}uszczak, Marta and Sch{\"a}fer, Wolfgang and Szczurek, Antoni},
    title = "{Probing gluon GTMDs of the proton in deep inelastic diffractive dijet production at HERA}",
    eprint = "2403.15110",
    archivePrefix = "arXiv",
    primaryClass = "hep-ph",
    doi = "10.1103/PhysRevD.110.054027",
    journal = "Phys. Rev. D",
    volume = "110",
    number = "5",
    pages = "054027",
    year = "2024"
}

@article{Boer:2023mip,
    author = {Boer, Dani{\"e}l and Setyadi, Chalis},
    title = "{Probing gluon GTMDs through exclusive coherent diffractive processes}",
    eprint = "2301.07980",
    archivePrefix = "arXiv",
    primaryClass = "hep-ph",
    doi = "10.1140/epjc/s10052-023-12040-6",
    journal = "Eur. Phys. J. C",
    volume = "83",
    number = "10",
    pages = "890",
    year = "2023"
}

@article{Boer:2021upt,
    author = {Boer, Dani{\"e}l and Setyadi, Chalis},
    title = "{GTMD model predictions for diffractive dijet production at EIC}",
    eprint = "2106.15148",
    archivePrefix = "arXiv",
    primaryClass = "hep-ph",
    doi = "10.1103/PhysRevD.104.074006",
    journal = "Phys. Rev. D",
    volume = "104",
    number = "7",
    pages = "074006",
    year = "2021"
}

@article{Ji:2016jgn,
    author = "Ji, Xiangdong and Yuan, Feng and Zhao, Yong",
    title = "{Hunting the Gluon Orbital Angular Momentum at the Electron-Ion Collider}",
    eprint = "1612.02438",
    archivePrefix = "arXiv",
    primaryClass = "hep-ph",
    doi = "10.1103/PhysRevLett.118.192004",
    journal = "Phys. Rev. Lett.",
    volume = "118",
    number = "19",
    pages = "192004",
    year = "2017"
}

@article{Hatta:2016dxp,
    author = "Hatta, Yoshitaka and Xiao, Bo-Wen and Yuan, Feng",
    title = "{Probing the Small- x Gluon Tomography in Correlated Hard Diffractive Dijet Production in Deep Inelastic Scattering}",
    eprint = "1601.01585",
    archivePrefix = "arXiv",
    primaryClass = "hep-ph",
    reportNumber = "YITP-16-1",
    doi = "10.1103/PhysRevLett.116.202301",
    journal = "Phys. Rev. Lett.",
    volume = "116",
    number = "20",
    pages = "202301",
    year = "2016"
}

@article{Bhattacharya:2022vvo,
    author = "Bhattacharya, Shohini and Boussarie, Renaud and Hatta, Yoshitaka",
    title = "{Signature of the Gluon Orbital Angular Momentum}",
    eprint = "2201.08709",
    archivePrefix = "arXiv",
    primaryClass = "hep-ph",
    doi = "10.1103/PhysRevLett.128.182002",
    journal = "Phys. Rev. Lett.",
    volume = "128",
    number = "18",
    pages = "182002",
    year = "2022"
}

@article{Bhattacharya:2024sck,
    author = "Bhattacharya, Shohini and Boussarie, Renaud and Hatta, Yoshitaka",
    title = "{Exploring orbital angular momentum and spin-orbit correlations for gluons at the Electron-Ion Collider}",
    eprint = "2404.04209",
    archivePrefix = "arXiv",
    primaryClass = "hep-ph",
    doi = "10.1103/PhysRevD.111.034019",
    journal = "Phys. Rev. D",
    volume = "111",
    number = "3",
    pages = "034019",
    year = "2025"
}

@article{Chall:2026oes,
    author = {Chall, Trambak Jyoti and {\L}uszczak, Marta and Sch{\"a}fer, Wolfgang and Szczurek, Antoni},
    title = "{Probing GPDs in exclusive electroproduction of dijets}",
    eprint = "2603.09686",
    archivePrefix = "arXiv",
    primaryClass = "hep-ph",
    doi = "10.1103/p7m5-qqyx",
    journal = "Phys. Rev. D",
    volume = "113",
    number = "11",
    pages = "114012",
    year = "2026"
}

@article{Berthou:2015oaw,
    author = "Berthou, B. and others",
    title = "{PARTONS: PARtonic Tomography Of Nucleon Software}: {A computing framework for the phenomenology of Generalized Parton Distributions}",
    eprint = "1512.06174",
    archivePrefix = "arXiv",
    primaryClass = "hep-ph",
    doi = "10.1140/epjc/s10052-018-5948-0",
    journal = "Eur. Phys. J. C",
    volume = "78",
    number = "6",
    pages = "478",
    year = "2018"
}

@article{Goloskokov:2006hr,
    author = "Goloskokov, S. V. and Kroll, P.",
    title = "{The Longitudinal cross-section of vector meson electroproduction}",
    eprint = "hep-ph/0611290",
    archivePrefix = "arXiv",
    reportNumber = "WU-B-06-02, WU B 06-02",
    doi = "10.1140/epjc/s10052-007-0228-4",
    journal = "Eur. Phys. J. C",
    volume = "50",
    pages = "829--842",
    year = "2007"
}

@article{Goloskokov:2007nt,
    author = "Goloskokov, S. V. and Kroll, P.",
    title = "{The Role of the quark and gluon GPDs in hard vector-meson electroproduction}",
    eprint = "0708.3569",
    archivePrefix = "arXiv",
    primaryClass = "hep-ph",
    reportNumber = "WU-B-07-07, WU B 07-07",
    doi = "10.1140/epjc/s10052-007-0466-5",
    journal = "Eur. Phys. J. C",
    volume = "53",
    pages = "367--384",
    year = "2008"
}

@article{Deja:2023ahc,
    author = "Deja, K. and Martinez-Fernandez, V. and Pire, B. and Sznajder, P. and Wagner, J.",
    title = "{Phenomenology of double deeply virtual Compton scattering in the era of new experiments}",
    eprint = "2303.13668",
    archivePrefix = "arXiv",
    primaryClass = "hep-ph",
    reportNumber = "CPHT-RR012.032022",
    doi = "10.1103/PhysRevD.107.094035",
    journal = "Phys. Rev. D",
    volume = "107",
    number = "9",
    pages = "094035",
    year = "2023"
}

@article{Guest:2016iqz,
    author = "Guest, Daniel and Collado, Julian and Baldi, Pierre and Hsu, Shih-Chieh and Urban, Gregor and Whiteson, Daniel",
    title = "{Jet Flavor Classification in High-Energy Physics with Deep Neural Networks}",
    eprint = "1607.08633",
    archivePrefix = "arXiv",
    primaryClass = "hep-ex",
    doi = "10.1103/PhysRevD.94.112002",
    journal = "Phys. Rev. D",
    volume = "94",
    number = "11",
    pages = "112002",
    year = "2016"
}

@article{CMS:2017wtu,
    author = "Sirunyan, A. M. and others",
    collaboration = "CMS",
    title = "{Identification of heavy-flavour jets with the CMS detector in pp collisions at 13 TeV}",
    eprint = "1712.07158",
    archivePrefix = "arXiv",
    primaryClass = "physics.ins-det",
    reportNumber = "CMS-BTV-16-002, CERN-EP-2017-326",
    doi = "10.1088/1748-0221/13/05/P05011",
    journal = "JINST",
    volume = "13",
    number = "05",
    pages = "P05011",
    year = "2018"
}

@article{Qu:2019gqs,
    author = "Qu, Huilin and Gouskos, Loukas",
    title = "{ParticleNet: Jet Tagging via Particle Clouds}",
    eprint = "1902.08570",
    archivePrefix = "arXiv",
    primaryClass = "hep-ph",
    doi = "10.1103/PhysRevD.101.056019",
    journal = "Phys. Rev. D",
    volume = "101",
    number = "5",
    pages = "056019",
    year = "2020"
}

@article{Dudek:2012vr,
    author = "Dudek, Jozef and others",
    title = "{Physics Opportunities with the 12 GeV Upgrade at Jefferson Lab}",
    eprint = "1208.1244",
    archivePrefix = "arXiv",
    primaryClass = "hep-ex",
    reportNumber = "JLAB-PHY-12-1599",
    doi = "10.1140/epja/i2012-12187-1",
    journal = "Eur. Phys. J. A",
    volume = "48",
    pages = "187",
    year = "2012"
}

@article{Moutarde:2018kwr,
    author = "Moutarde, H. and Sznajder, P. and Wagner, J.",
    title = "{Border and skewness functions from a leading order fit to DVCS data}",
    eprint = "1807.07620",
    archivePrefix = "arXiv",
    primaryClass = "hep-ph",
    doi = "10.1140/epjc/s10052-018-6359-y",
    journal = "Eur. Phys. J. C",
    volume = "78",
    number = "11",
    pages = "890",
    year = "2018"
}

@article{Kroll:2012sm,
    author = "Kroll, Peter and Moutarde, Herve and Sabatie, Franck",
    title = "{From hard exclusive meson electroproduction to deeply virtual Compton scattering}",
    eprint = "1210.6975",
    archivePrefix = "arXiv",
    primaryClass = "hep-ph",
    reportNumber = "IRFU-12-174, WUB-12-22, IRFU-12-174; WUB/12-22",
    doi = "10.1140/epjc/s10052-013-2278-0",
    journal = "Eur. Phys. J. C",
    volume = "73",
    number = "1",
    pages = "2278",
    year = "2013"
}

@article{Kumericki:2009uq,
    author = "Kumeri{\v{c}}ki, Kresimir and Mueller, Dieter",
    title = "{Deeply virtual Compton scattering at small $x_B$ and the access to the GPD H}",
    eprint = "0904.0458",
    archivePrefix = "arXiv",
    primaryClass = "hep-ph",
    doi = "10.1016/j.nuclphysb.2010.07.015",
    journal = "Nucl. Phys. B",
    volume = "841",
    pages = "1--58",
    year = "2010"
}

@article{Cuic:2020iwt,
    author = {{\v{C}}ui{\'c}, Marija and Kumeri{\v{c}}ki, Kre{\v{s}}imir and Sch{\"a}fer, Andreas},
    title = "{Separation of Quark Flavors Using Deeply Virtual Compton Scattering Data}",
    eprint = "2007.00029",
    archivePrefix = "arXiv",
    primaryClass = "hep-ph",
    reportNumber = "ZTF-EP-20-04",
    doi = "10.1103/PhysRevLett.125.232005",
    journal = "Phys. Rev. Lett.",
    volume = "125",
    number = "23",
    pages = "232005",
    year = "2020"
}

@article{Kumericki:2016ehc,
    author = "Kumericki, Kresimir and Liuti, Simonetta and Moutarde, Herve",
    title = "{GPD phenomenology and DVCS fitting}: {Entering the high-precision era}",
    eprint = "1602.02763",
    archivePrefix = "arXiv",
    primaryClass = "hep-ph",
    doi = "10.1140/epja/i2016-16157-3",
    journal = "Eur. Phys. J. A",
    volume = "52",
    number = "6",
    pages = "157",
    year = "2016"
}

@article{Moutarde:2019tqa,
    author = "Moutarde, H. and Sznajder, P. and Wagner, J.",
    title = "{Unbiased determination of DVCS Compton Form Factors}",
    eprint = "1905.02089",
    archivePrefix = "arXiv",
    primaryClass = "hep-ph",
    doi = "10.1140/epjc/s10052-019-7117-5",
    journal = "Eur. Phys. J. C",
    volume = "79",
    number = "7",
    pages = "614",
    year = "2019"
}

@article{Cichy:2024afd,
    author = "Cichy, Krzysztof and Constantinou, Martha and Sznajder, Pawe{\l} and Wagner, Jakub",
    title = "{Nucleon tomography and total angular momentum of valence quarks from synergy between lattice QCD and elastic scattering data}",
    eprint = "2409.17955",
    archivePrefix = "arXiv",
    primaryClass = "hep-ph",
    doi = "10.1103/PhysRevD.110.114025",
    journal = "Phys. Rev. D",
    volume = "110",
    number = "11",
    pages = "114025",
    year = "2024"
}

@article{Mankiewicz:1997aa,
    author = "Mankiewicz, L. and Piller, G. and Weigl, T.",
    title = "{Hard leptoproduction of charged vector mesons}",
    eprint = "hep-ph/9712508",
    archivePrefix = "arXiv",
    reportNumber = "TUM-T39-97-33",
    doi = "10.1103/PhysRevD.59.017501",
    journal = "Phys. Rev. D",
    volume = "59",
    pages = "017501",
    year = "1999"
}

@article{Duplancic:2022ffo,
    author = "Duplan{\v{c}}i{\'c}, Goran and Nabeebaccus, Saad and Passek-Kumeri{\v{c}}ki, Kornelija and Pire, Bernard and Szymanowski, Lech and Wallon, Samuel",
    title = "{Accessing chiral-even quark generalised parton distributions in the exclusive photoproduction of a $ \gamma \pi ^{\pm} $ pair with large invariant mass in both fixed-target and collider experiments}",
    eprint = "2212.00655",
    archivePrefix = "arXiv",
    primaryClass = "hep-ph",
    doi = "10.1007/JHEP03(2023)241",
    journal = "JHEP",
    volume = "03",
    pages = "241",
    year = "2023"
}

@article{Bellotti:2026bau,
    author = {Bellotti, Mattia and Boer, Dani{\"e}l and Pisano, Cristian},
    title = "{Gluon GTMDs in the exclusive electroproduction of heavy-quark pairs}",
    eprint = "2607.14964",
    archivePrefix = "arXiv",
    primaryClass = "hep-ph",
    journal = "",
    year = "2026"
}
\end{document}